\documentclass[extra]{gji}
\usepackage{timet,color}
\usepackage{float}
\usepackage{graphicx}
\usepackage{amsmath}
\usepackage[urlcolor=blue,citecolor=black,linkcolor=black]{hyperref}

\title[Impacts in Granular media]
  {
  Elastic waves generated by impact and vibration in confined granular media
  }

\begin{document}

\author[Gallot et. al.]
  {T. Gallot$^1$, C. Sedofeito$^1$, A. Ginares, G. Tancredi$^1$ \\
  $^1$ , Instituto de F\'isica, Facultad de Ciencias, Universidad de la Rep\'ublica, Montevideo, Uruguay.
  }

\let\leqslant=\leq

\label{firstpage}

\maketitle

\maketitle
\begin{summary}
Observational data of asteroids can be explained by considering them as an agglomerate of granular material. Understanding the mechanical properties of these objects is relevant for many scientific reasons: space missions design, evaluation of impact threats to our planet, and understanding the nature of asteroids and their implication in the origin of the solar system. In-situ measurements of mechanical properties require complex and costly space missions. Here a laboratory-scale characterization of wave propagation in granular media is presented using a novel experimental setup as well as numerical simulations. The pressure inside an asteroid is still a matter of debate, but it definitely presents a pressure gradient towards the interior. This is why impact characterization needs to be performed as a function of the confining pressure. Our experimental setup allows for the simultaneous measurement of the external confining pressure, internal pressure, total strain, and acceleration in a $50$ cm side squared box filled up with a billion grains. We study the propagation of impact-generated and shaker-born seismic body waves in the 500 Hz range. Through subsequent compression-relaxation cycles, it was observed that the granular media behaves on average like a solid with a constant elastic modulus during each compression. Effective medium theory (EMT) for granular media explains the data at low pressure. After each compression-relaxation cycle, the elastic modulus increases, and a high hysteresis is observed: relaxation shows a more complex behavior than compression. We show that seismic waves generated by   both impact and vibration travels at the pressure wave speed. Thanks to a numerical model, we measure a strong wave attenuation $\alpha\sim3.4$ Np/m. We found that the wave speed increases with the confining pressure with a $p^{1/2}$ dependency, in disagreement with theoretical models that predicts a shallower dependency. The dependency of the elasticity with the confining pressure can be explained by a modified EMT model with a coordination number proportional to the pressure, or equivalently by a mesoscopic nonlinear model based on third-order nonlinear elastic energy. The interpretation of these models is a deep reorganization in the particle contact network.

\end{summary}
\begin{keywords}
    PHYSICAL PROPERTIES: Elasticity and anelasticity
    GEOGRAPHICAL: Extraterrestrial
    SEISMOLOGY: Acoustic properties, Wave propagation, Body waves
\end{keywords}

\section{Introduction}

Granular media are often used as laboratory-scale systems for complex natural phenomena such as seismic fault gauges \citep{Planet2015}. There are clear observations that asteroids are agglomerates of rocks like the rubble-pile asteroid, Itokawa, observed by Hyabusa \citep{fujiwara2006rubble}; as well as many other observational pieces of evidences like the rotational spin-barrier on asteroids larger than a few hundred meters and the crater chains observed in the surface of the Galilean satellites (see \textit{e.g.} \citet{walsh2018}. This work is thus motivated by the need for experimental data to understand how asteroids respond to impacts. Understanding the nature of asteroids is relevant for earth collision hazard assessment and asteroids exploration \citep{hestroffer2019small}. It may also be important to comprehend the collisional processes in the formation and evolution of our solar system  \citep{holsapple1993scaling}. More particularly, we are also interested in understanding the nature of the so-called active asteroids \citep{jewitt2012active}; asteroids that show a temporary tail, that could be generated by a shaken mechanism induced by the propagation of  seismic waves into the interior \citep{tancredi2015main,tancredi2022}. 

One of the alternatives to deflect an asteroid on course to collide with the Earth is kinetic impact: hitting the body with a massive object to transfer a linear impulse that changes its course. NASA launched the DART mission to test this technology; the experiment successfully occurred on Sept. 26, 2022 \citep{rivkin2021}. The efficiency of this process depends on the cratering event \citep{stickle2022}, the propagation of the impact-induced seismic wave into the interior of the body \citep{tancredi2022}, and the ejecta distribution \citep{fahnestock2022}. The images released at the time of impact showed that the target, the 160m asteroid Dimorphos, resembles a rubble-pile. With the NASA-DART mission, impacting an asteroid to deflect its trajectory is not science fiction anymore. The Impact creates an important material ejection observed seconds after the impact, but the brightness increase is still measurable two weeks after \footnote{See NASA News: https://www.nasa.gov/feature/nasa-dart-imagery-shows-changed-orbit-of-target-asteroid}. This means that materials escape at very low velocity, and that the direct impact is not the only reason for ejection, seismic waves propagating from the impact all around the asteroid are also responsible for the material ejection \citep{tancredi2022}.

A few experimental works study impacts on grains in unconfined \citep{yasui2015experimental} or confined \citep{van2013shock,martinez2021extending} media. The micro-gravity on asteroids confers mechanical properties to granular media that are not easy to reproduce on earth \citep{altshuler2014settling,villalobos2022geometrical}. However, these experiments are fundamental to validate numerical models, in particular Discrete Element Models (DEM)  \citep{schopfer2009impact, wang2009esys_particle} for the complex physics of granular mechanics \citep{Duran2012}.

Inhomogeneity of grain packing \citep{Liu1995,jaeger1996granular}, together with material relaxation \citep{alexander1998amorphous}, explain most  of the complexity in granular media. The contacts between the grains form a network that reorganizes under stresses \citep{mueth1998force,howell1999stress,Cambau2013}. 
Because of this reorganization, most of the numerical and laboratory experiments on grains begin with the preparation of the material. The quasi-static problem of stress distribution has been addressed by Janssen model \citep{nedderman1992statics}. The observation of the chain force provides some insight to discuss the model limitation due to correlation length, microscopic features, reorganization, and hysteresis \citep{Ovarlez2003,Ovarlez2005,Gennes1999}.

A spectacular consequence of reorganization is a jamming transition from fluid to solid state \citep{cates1999jamming,liu1998jamming,van2009jamming}. Wave propagation in grains is an amazing probing tool for mechanical parameters in granular media \citep{Somfai_2005,jacob2008acoustic,silbert2005vibrations}, but its understanding is still challenging \citep{luding2005information}, because of a variety of phenomena such as nonlinear propagation \citep{zhang2020nonlinear}, nonlinear constitutive equations \citep{Renaud_2013,goddard1990nonlinear,trarieux_2014}, wave dispersion \citep{Chrzaszcz2016waves,cheng2020elastic}, multiple scattering \citep{Jia_2004,tell2020acoustic,langlois2015,trujillo2011multiple,brunet2006etude,Page_1996}, or path-dependent propagation \citep{hua2019wave,owens2011sound}.

Effective Medium Theory (EMT) \citep{walton1987effective} predicts a scaling of the coherent wave speeds with pressure between $p^{1/6}$
for Hertzian contact, or $p^{1/3}$ considering non-Hertzian contact or variation in the coordination number $C$ \citep{goddard1990nonlinear}.
Discrete Element Models (DEM) and experimental observations confirmed  these numbers \citep[see ][for a non-exhaustive review]{jia2021elastic}.
The contact between two grains can be described by Hertz-like models; then, EMT stipulates that the macroscopic response of a medium is the sum of an
averaged grain-grain contact \citep{ovarlez2005elastic,kocharyan2022influence}. This strong hypothesis of linearity explain why EMT fail to explains
many observations where the scaling law exponent is shown to depend on the pressure range \citep{makse2004granular}, stress history \citep{cheng2020elastic},
wave macroscopic amplitude \citep{wichtmann2004influence}, and local amplitude around force chains \citep{owens2011sound}.

In this work we propose an experimental study of laboratory scaled asteroid impacts. We use granular media as a model asteroid. There are two fundamental differences between a real asteroid and our experiment: the gravity conditions and the impact velocity. Self gravity induces a pressure distribution inside an asteroid that is not well known \citep[see different estimates by: ][]{cheng2004,sharma2013,zhang2018}; but it certainly presents a pressure gradient with increasing values towards the interior of the body. For this reason, the granular media is confined and the impacts are realized for different confining pressure steps. As regard to the low velocity of our impactors, we did study neither the crater geometry nor the energy transfer that would depend on the impact velocity. Instead we were interested in the wave propagation outside of the impact zone.

In the present experimental work, we face the whole complexity of the quasi-static and dynamic mechanical response of granular media. 
In Section \ref{sec:Experimental-set-up} we present the experimental setup, the characteristic of the materials and the devices used in the experiments.
This is why our main parameter is confining pressure of the granular media. In Section \ref{sec:Quasi-static-characterization} we  present the quasi-static response in glass beads. Then, the results of the impact-generated and shaker-born seismic waves as a function of the confining pressure are presented in Section \ref{sec:Dynamic-characterization}.

\section{Experimental setup}\label{sec:Experimental-set-up}

\begin{figure}
\centering
\includegraphics[width=0.95\columnwidth]{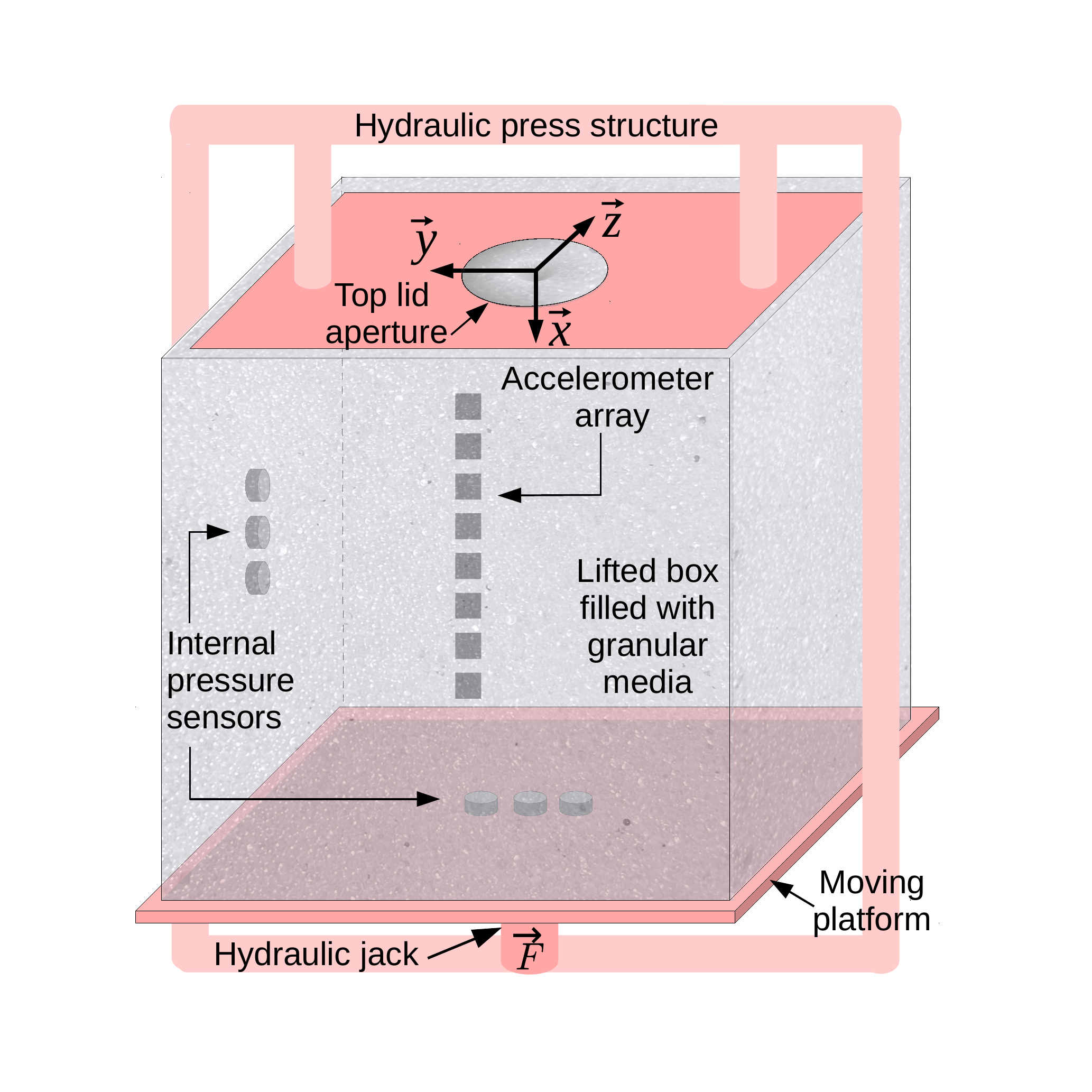}

\caption{\label{fig:scheme}{A 50-cm side acrylic cubic box filled with granular material is set on a moving platform. The granular media is confined inside the box lifted by a hydraulic jack while the top cover, welded on the structure, stays unmoved. The top lid aperture allows direct contact between the granular media and  the projectile or shaker. Internal pressure is monitored with six sensors placed half on a lateral wall and half on the floor of the box. The vibrations generated at the aperture are registered by a vertical array of 3-axis accelerometers immersed in the media, at a horizontal distance of 9 cm from the centre of the lid aperture.}}
\end{figure}

The experimental approach focuses on the propagation of short waves generated by perturbations due to impacts or vibration on the surface of a box containing a confined granular media. The box is a cube of side $L=50$ cm (internal distance between the lateral walls, Fig. \ref{fig:scheme}). The walls are made of 14 mm thick transparent acrylic. The cube rests on a moving platform with a sliding top lid. A circular opening of 16 cm in diameter allows the direct impact of the projectile or the contact of the shaker with the material. The inner top lid is stationary as it is welded to the hydraulic press structure. The box is uplifted by the hydraulic jack (Enerpac RC106 with a 15-cm stroke), compressing the material. The hydraulic press has been designed for a 10 tonnes maximum load. 

\begin{figure}
\centering
\includegraphics[width=0.95\columnwidth]{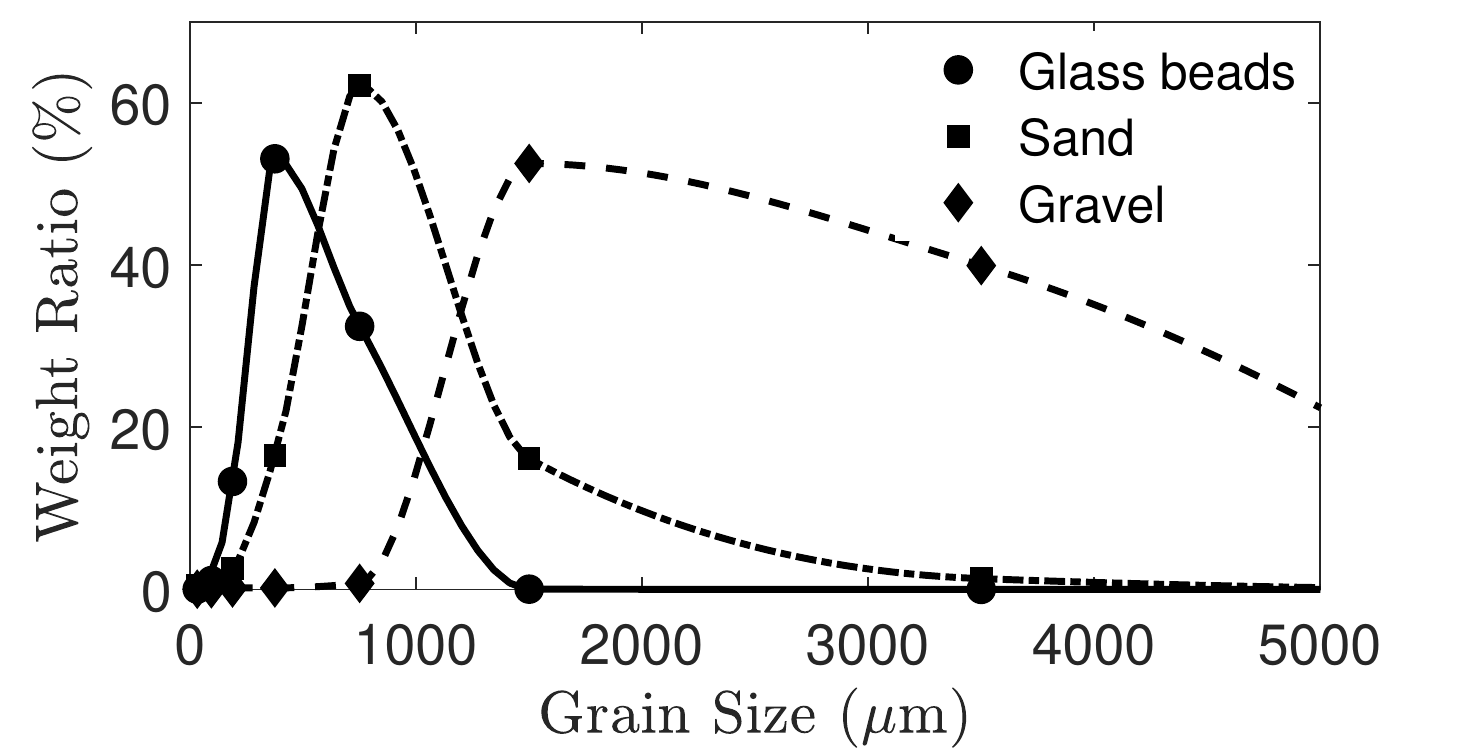}\caption{\label{fig:Grain size}Size distribution of (1) Sand with quartz-feldspathic
composition. (2) Gravel of mainly lithic composition with mostly
granite clasts. (3) Glass beads.}
\end{figure}

Experiments are performed using three different granular materials: glass beads (artificial), sand, and gravel (both natural). Size distributions are shown in Fig. \ref{fig:Grain size}, while angularity,  sphericity, density, and volume fraction are described in Table \ref{table:grains}. The granular material, stored in a 100-liter barrel, is positioned over the hydraulic press structure using an electric winch. A plug
at the bottom of the barrel  is removed releasing the material, filling up the box through its upper aperture.  By the end of this process, the accelerometers and pressure sensors inside the box are completely covered (see below for a description of the location of these devices). The barrel is weighted in order to have 195$\pm1$ kg of grains inside the box. After discharge, the pile needs to be
manually even. Material preparation consists of a series of five compression-relaxation cycles from 0 to 5 tonnes. This procedure rearranges the grains on the top of the pile, flattening the surface.

\begin{table}
\centering
\caption{\label{table:grains}  Characteristics of the granular media used in the experiments: (1) Glass beads, with zero angularity and a high spherical shape ratio. (2) Sand grains are angular to sub-angular and have a shape ratio of medium sphericity. (3) Gravel grains are angular and of very low sphericity. The Diameter is the mode of the size distribution of Fig. \ref{fig:Grain size}.}

\begin{tabular}{|c|c|c|c|c|} 
\hline 
Material & Diameter & Grain size & Density & Vol. frac. \tabularnewline
        & d($\mu m$) & (Wentworth scale) &$\rho$ (g/ml)    &  $\phi$ \tabularnewline
\hline 
\hline 
Glass beads & 250 & fine to medium sand& 1.63& 0.66  \tabularnewline
\hline 
Sand & 500 & coarse sand &  1.66 & 0.64 \tabularnewline
\hline 
Gravel & 1500 & medium to coarse gravel & 1.66 & 0.63  \tabularnewline
\hline 
\end{tabular}

\end{table}

The coordinate system ($\overrightarrow{x}$,$\overrightarrow{y}$,$\overrightarrow{z}$)
has its origin at the impact point as shown in Fig. \ref{fig:scheme}. The perturbations are generated along the $x$-direction. An array of 3-axis accelerometers (Analog Devices, ADXL327, $\pm2$g sensitivity 0.42 V/g) embedded in the granular media registers the vibrations.  The accelerometers were located in a vertical array, at a horizontal distance of $y_0=9$ cm from the impact zone, to prevent the destruction of the devices by the penetrating bullet. Hydraulic jack pressure is measured using a pressure transmitter (Wika A-10). We deduce the hydraulic force  $\overrightarrow{F}$  by considering the cylinder effective area (manufacturer data 14.5 $\textrm{cm}^{2}$).
Also, piezo-resistive gauge pressure sensors (LEEG, LG190H704G) are positioned along the walls, in direct contact with the granular material, to measure internal stresses.  The sensors have a 18-mm diameter circular active area, much larger than the grain sizes. The sensitivity of each gauge is calibrated using a 3 m water column to check repeatability and linearity.

The box displacement is monitored with a digital camera with a 1-s time-lapse (Pixelink PL-D722). Assuming a displacement in the $\overrightarrow{x}$ direction, the gray-scaled image can be averaged along the $\overrightarrow{y}$-axis. The correlation between the first image and all the following is then computed. The position of  maximum correlation gives an estimation of the box displacement, $u_{x}$,  with a 50-$\mu$m uncertainty. Particle Image Velocimetry trials were performed, but no box deformation nor relative displacement of the
grains could be measured.

The experiment requires measuring the following physical parameters: (1) confining pressure, (2) internal pressure, (3) box displacement, and (4) acceleration of seismic waves, all at the same time. The acquisition of these quantities is performed by the aforementioned devices: (1) pressure sensor in the hydraulic jack piston; (2) six pressure transmitters placed on the walls of the box (represented by gray cylinders in Fig. \ref{fig:scheme}); (3) a digital camera; and (4) 3D accelerometers array embedded in the media positioned every 3 cm from $x=10$ cm to $x=37$ cm, at $y=9$ cm and $z=0$ (represented by gray squares in Fig. \ref{fig:scheme}). The acquisition is performed by two digitizer cards (National Instrument USB-6010, 250 kHz, 16 channels) controlled by Matlab. Reading the internal PC clock is needed to synchronize two Matlab sessions running in parallel to control each card; one for the pressure, and one for acceleration. Additionally, one of the cards switches on a led for camera synchronization.

Two different experiments are performed on each granular material. In the first one, waves are generated by projectiles impacting the media. In the second one, waves are generated by a shaker in contact with the media through the top lid aperture (see Fig. \ref{fig:scheme}). For impacts, projectile shots are triggered manually. Adequate ear and eye protection was used. Verbal coordination between two operators was needed to capture the impact within a 5-s acquisition of the sensors. Three devices were used to accelerate spherical projectiles: a spring-piston air rifle, a $CO_2$ pistol, and a crossbow. The bullets and guns are described in Table \ref{table:guns}.  A function generator sends an input signals for the shaker and triggers the acquisition cards for synchronization.

\begin{table}
\centering
\caption{\label{table:guns} Characteristics of the shooting devices and their bullets.
Guns: spring-piston air rifle, $CO_2$ pistol, and crossbow. For each gun, we listed the characteristics of the corresponding projectile: diameter, mass and material. The velocity was measured with a bullet chronograph (PosChrono DLX).  The uncertainty was computed as the standard deviation over 35 shoots. The energy column refers to the kinetic energy $e=mv^2/2$, with 1\% uncertainty omitted in the table.}

\begin{tabular}[h!]{|c|c|c|c|c|c|} 
\hline 
 Device & \multicolumn{5}{c}{Projectile properties} \tabularnewline
\hline 
\hline 
Gun & Diameter & Mass & Material & Velocity & Energy \tabularnewline
&  (mm)  & $m$ (g) &  & $v$ (m/s)  &  $e$ (kJ) \tabularnewline
\hline 
Riffle & 5.5 & 1.0  & lead & $239\pm4$ & 28.6 \tabularnewline
\hline 
 Pistol & 4.5 & 0.35 & steel & $171\pm3$ & 5.1 \tabularnewline
\hline 
 Crossbow & 6 & 0.9 & copper & $66\pm2$ & 1.9\ \tabularnewline
\hline 
\end{tabular}

\end{table}

\section{Quasi-static characterization}\label{sec:Quasi-static-characterization}

Since dynamic parameters are to be measured as a function of the
confining pressure, the distribution of stress inside the granular
media needed particular attention. In this section, we neglect the effects of the aperture on the top lid, and the friction on the side walls. This assumption is justified by the three orders of magnitude between the grains and the box sizes. Under these odeometric conditions, the granular media only experiences external
compressive stress from the six side walls. We adopt the classical stress tensor notation in a solid \citep{landau1986theory}, where the compressive stresses are positive.

\begin{figure}
\centering \includegraphics[width=0.95\columnwidth]{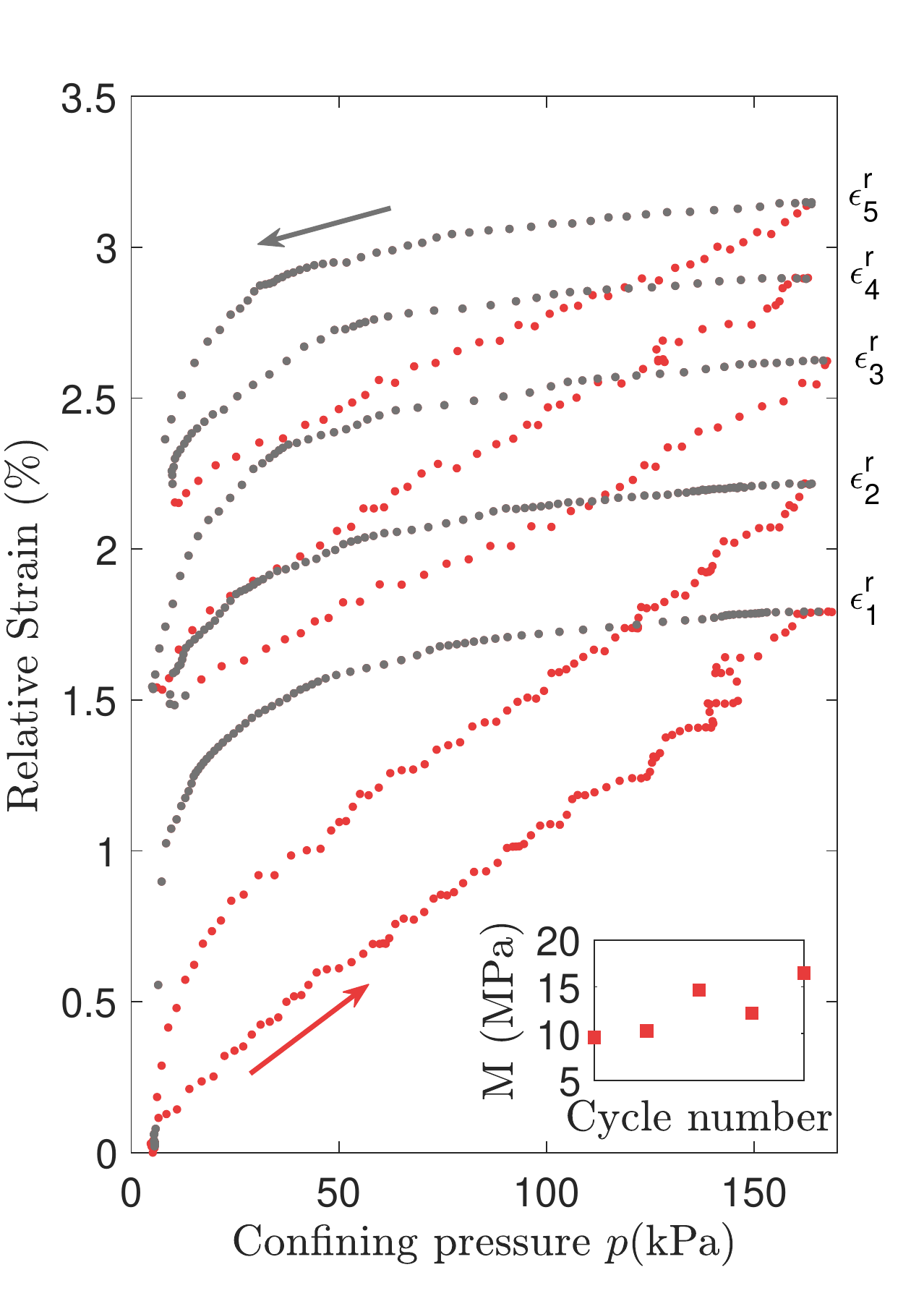}
\caption{Evolution of the strain $\epsilon$ as a function of the confining
pressure $p$ during five oedometric cycles. Compression episodes exhibit proportionality between $\epsilon$ and $p$. The inverse of the slope
is the elastic modulus $M$, represented as red squares in the inset.
Pressure decreases much faster than strain during relaxation episodes
in gray. After five cycles of compression-relaxation, the final strain
is 3\% corresponding to a 15 mm displacement of the box. Compaction
is pictured by the maximum strain $\epsilon_{i}^{r}$ for each cycle
$i$. The increase of the elastic modulus is a consequence of compaction.}
\label{fig:ssRaw} 
\end{figure}

The confining pressure is controlled by the hydraulic force $\overrightarrow{F}$
applied on the bottom wall. It is 
defined as $p=\parallel\overrightarrow{F}\parallel/S$, with the contact
area $S=0.25$ m$^{2}$. The confining pressure is not a volumetric pressure but a macroscopic stress. As a granular media presents heterogeneities,
the quantities measured locally are named $Q$ while the effective
measurements on the whole confined medium are named $\overline{Q}$. Fig. \ref{fig:ssRaw} shows the relative vertical strain $\overline{\epsilon_{xx}}=\epsilon$ 
as a function of the confining pressure. Five compression-relaxation
cycles were recorded. The minimum pressure $p_{min}=7.6$ kPa corresponds
to the material weight distributed over the box floor area. The maximum
pressure is set manually at $p_{max}=164\pm5$ kPa. To picture hysteresis,
the pale red dots correspond to the compression phase; while the gray
dots correspond to the relaxation phase. 

\subsection{Granular media as a quasi-elastic solid}

During compression, strain increases linearly above 25 kPa. Since
the system is considered to be non-dissipative, confining pressure and effective stress are the same throughout the box: $\overline{\sigma_{xx}}=p$.
The slope corresponds to the inverse of the apparent elastic modulus $M$:
\begin{equation}
p=M\epsilon\label{eq:Elastic modulus}
\end{equation}
This linear stress-strain relationship corresponds to a macroscopically homogeneous linear elastic media with no horizontal stress ($\overline{\epsilon_{yy}}=\overline{\epsilon_{zz}}=0$).
The elastic modulus is given by  $M=\lambda+2\mu$, where 
$\lambda$ and $\mu$ are the Lame's parameters (see Appendix \ref{sec:Apendicequasielasticity}, ec. \ref{eq:3}). The inset of Fig. \ref{fig:ssRaw} shows the variation of the elastic
modulus over five cycles (square dots). An increase from 10 MPa
to 20 MPa is visible due to the compaction of the media at the macro-scale after each cycle: the granular media is a different elastic media after each compaction.

After each compression-relaxation cycle $i$, the strain does not return to zero. We observed a small material leaking through the edges of the sliding top lid.
If the total strain $\epsilon\sim3\%$ would correspond to a material leaking, a mass variation $dm=\epsilon m \approx 5$kg should be observed. Instead, the grains recollected outside the box after a few cycles weights less than 10 gr and leaking effect can be neglected. Therefore, we conclude that the granular media experience a new compaction state because of material reorganization. 

\begin{figure}
\centering \includegraphics[width=0.95\columnwidth]{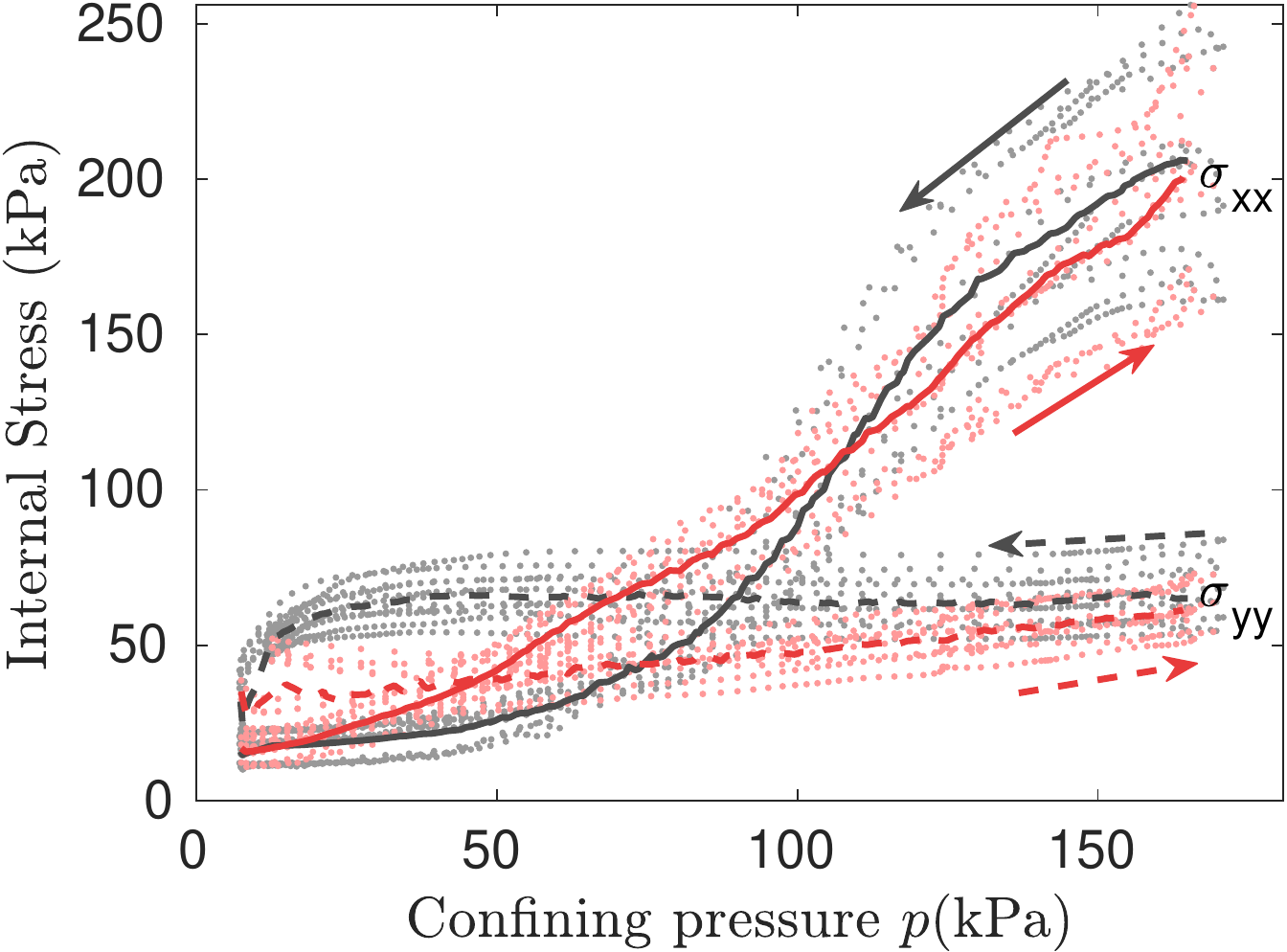}
\caption{Compressive stresses averaged on three sensors in glass beads are
shown to be proportional to the confining pressure $p$ applied by
the hydraulic press during compression (red straight line for $\sigma_{xx}$
and red dashed line for $\sigma_{yy}$). Relaxation in gray shows
a \textit{s-like} shape for $\sigma_{xx}$ and a constant $\sigma_{yy}$
until collapsing around 25 kPa. Pale dots represent raw data for the
six sensors with the same color code.\label{fig:Compressive-stresses-averaged}}
\end{figure}

Figure \ref{fig:Compressive-stresses-averaged} shows internal stresses
as a function of confining pressure $p$. $\sigma_{xx}$ is measured
by three sensors oriented along the $\overrightarrow{x}$-negative
direction (facing upwards) at the bottom of the box, $x$=50 cm,
centered in the $\overrightarrow{z}-$direction, $z=0$ cm, and at
$y$=\{-3, 0, 3\} cm (see Fig. \ref{fig:scheme}). Similarly, $\sigma_{yy}$ is measured with three sensors attached to the left side wall, $y=25$
cm, oriented along the $y$-negative direction and facing the grains.
The sensors were located at depths of $x$=\{22, 25, 28\} cm.

All stresses are plotted in red during compression and gray during
relaxation. The pale dots represent raw data from the six sensors
over five compression-relaxation cycles.  The average
of $\sigma_{xx}$ for each confining pressure over the five cycles and for
the three sensors, $<\sigma_{xx}>$, is represented as thick full lines. The same for $<\sigma_{yy}>$ represented in thick dashed lines.

The stress measured by the sensors showed a significant spatial variability;
around 30\% for $\sigma_{xx}$ and 20\% for $\sigma_{yy}$.
The variability between each cycle is under 5\% at any confining pressure.
Average internal stress increases proportionally with the confining
pressure only during compression. During relaxation, we observe a
\textit{s}-\textit{shape} for $<\sigma_{xx}>$, and a plateau collapsing
at low $p$ for $<\sigma_{yy}>$. Above $p=100$ kPa, all stresses
are larger during relaxation than compared to compression. This means
that elastic energy is stored at the beginning of the relaxation and
released when the confining pressure decreases under $p=100$ kPa.

Linear regressions for the apparent-linear portions  ($p>$50 kPa) have coefficients
of determination (R$^{2}$) of 0.99 for $<\sigma_{xx}^{c}>$ and 0.96
for $<\sigma_{yy}^{c}>$, where the \textit{$c$} superscript stands for stress during compression.
\begin{equation}
<\sigma_{xx}^{c}>=1.4p , \label{eq:sigmaxx}
\end{equation}
\begin{equation}
<\sigma_{yy}^{c}>=0.2p .\label{eq:sigmayy}
\end{equation}
A linear coefficient higher than 1 in eq. (\ref{eq:sigmaxx}) can be
explained by: (1) the limited area of measurement (around a thousand
grains) together with the high spatial variability observed in other
experiments \citep{erikson2002force,mueth1998force}; and (2) the high
probability of finding vertical forces higher than the average established experimentally \citep{Liu1995,Mueth1998}.

\cite{Evesque1998} introduced a quasi-elastic model for granular media where the stress redirection : 
\begin{equation}
\overline{\sigma_{yy}}=\overline{\sigma_{zz}}=K\overline{\sigma_{xx}},\label{eq:Poisson}
\end{equation}
gives a coefficient of redirection $K=\frac{\lambda}{\left(\lambda+2\mu\right)}$; (see Appendix \ref{sec:Apendicequasielasticity}
for details). Remembering that $p=\overline{\sigma_{xx}}$,  the redirection coefficient can be estimated experimentally as $K\equiv\frac{<\sigma_{yy}^{c}>}{p}=0.2$
(see eq. \ref{eq:sigmayy}). This corresponds to a Poisson ratio $\nu=\frac{K}{1+K}=0.16$ and is in agreement
with measurement in unconsolidated sands with $K$ ranging from 0.1 to 0.3 \citep{avseth2005seismic,spencer1994frame}.
The estimation of the redirection factor allows to compute the Lame's parameters (see Eq. \ref{eq:lambamu}). During compression in glass beads, we estimate an average $\lambda=2.4$ MPa and $\mu=4.8$ MPa.

Despite a high spatial variability of the internal stresses, the granular media behaves macroscopically as a quasi-elastic media during each compression and for $p>50$ kPa. After each compression-relaxation cycle, the elastic modulus increases due to compaction. We observed that the elastic energy is stored and released during the relaxation. 


\subsection{Effective granular media \label{subsec:Data interpretation}}

\begin{figure}
\centering \includegraphics[width=0.95\columnwidth]{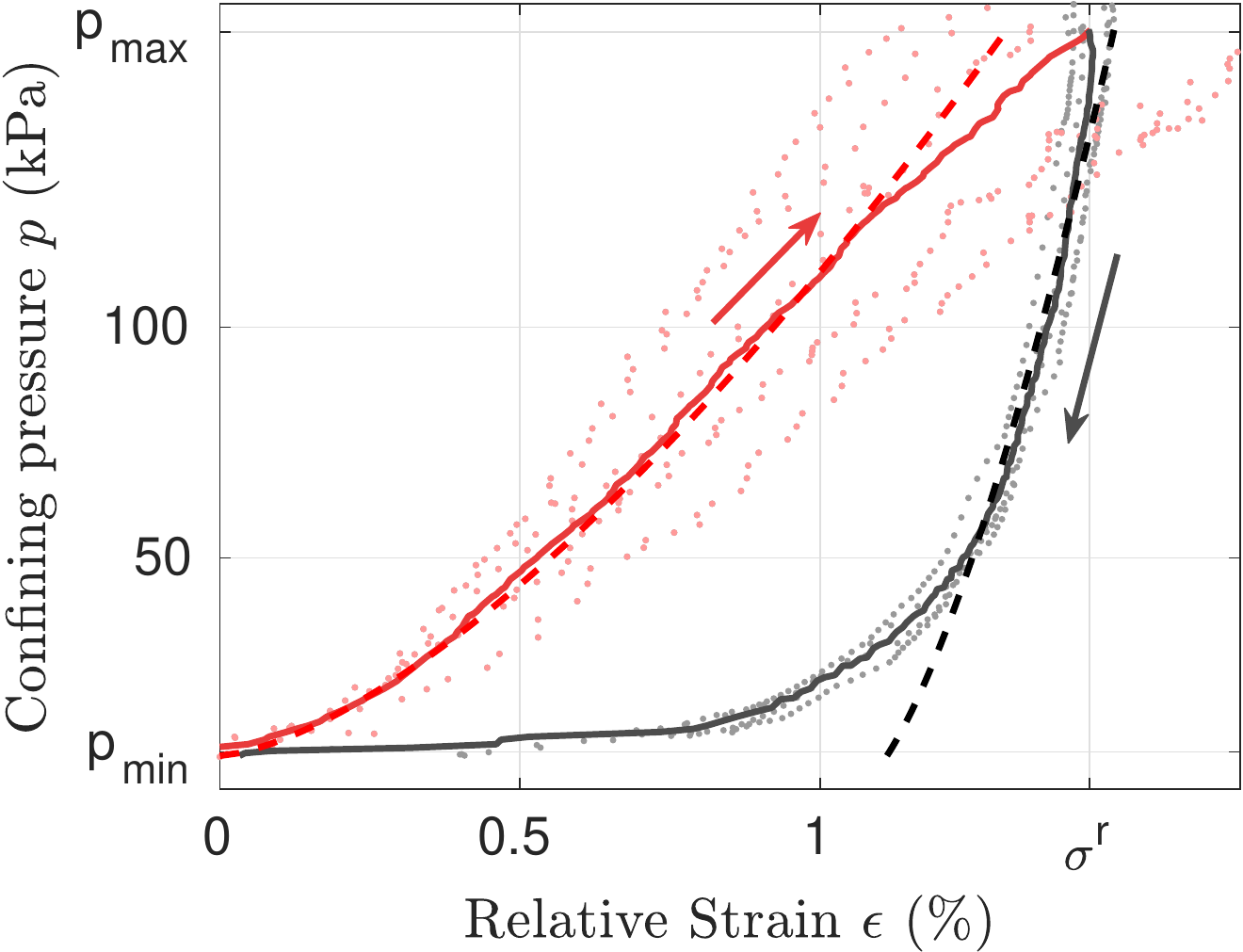}\caption{\label{StaticMeasurement} Confining pressure ( $p=\overline{\sigma_{xx}}$)
is represented as a function of axial strain ($\epsilon=\overline{\epsilon_{xx}}$)
over 5 oedometric cycles. Averaged data in full line shapes a closed
stress-strain curve. Compared to Fig. \ref{fig:ssRaw}, the non-reversible
compaction $\Delta_{i}$ is removed. Data are shown in red during
compaction and gray during relaxation. Raw data in pale dot shows
a high variability during compression and low variability during relaxation.
According to eq. (\ref{eq: EMT pres dep}), fitted curves are represented
in dashed lines with $S=580$ kPa in black, and $S=105$ kPa in red.}
\end{figure}

In this section, the effect of compaction is removed from the strain to help interpreting the data with the effective medium theory.  Compaction between cycles is quantified by $\Delta_{i}=\epsilon_{i+1}^{r}-\epsilon_{i}^{r}$,
where $\epsilon_{i}^{r}$ stands for the strain at the beginning of
each relaxation cycle $i$ (see Fig. \ref{fig:ssRaw}). In Fig.
\ref{StaticMeasurement}, the non-reversible compaction $\Delta_{i}$
is removed for each cycle $i+1$ during both compression and relaxation.
The strain at the beginning of the compression for each cycle is set
to zero. Similarly, the starting strain for the relaxation is set
at $\epsilon^{r}=1.7$ to close the hysteretic loop. The data points in each cycle are represented 
with small dots in red during compaction and gray during relaxation.
Averaged data represented
as a full line in Fig. \ref{StaticMeasurement} results in a closed
pressure-strain curve. This representation highlights that relaxation
follows the same path for each cycle and does not depend on compaction
stage. On the other hand, compression path changes after each compaction
stage.

EMT considers Hertz contact with average stress between each grain
\citep{makse1999effective}. A typical expression of the inter-grain
force can be summarized as: 
\begin{equation}
\overrightarrow{F}=\sqrt{\xi_{n}R_{eff}}\left[k_{n}\xi_{n}\overrightarrow{n}+k_{s}\xi_{s}\overrightarrow{s}\right],\label{eq:InterGrainForce}
\end{equation}
where $k_n,k_s$ are the normal and shear elastic stiffness, respectively, $\xi_n,\xi_s$
are the normal, and shear displacements between grains, and $R_{eff}$
is the effective radius. According to such a contact law, the EMT predicts the following relationship between pressure and strain (\citet{roux2015pre} see Appendix \ref{sec:Apendice-B-EMT}):
\begin{equation}
p=S_{e}\epsilon^{3/2},\label{eq: EMT pres dep}
\end{equation}
where $S_{e}$ is the oedometric effective stiffness. Stress-strain curves in Fig. \ref{StaticMeasurement} are fitted using this expression. During compression, the best fit for the average data gives $S_{e}=105$ MPa. Compaction
explains why $S_{e}$ is ranging from 77 MPa to 152 MPa for the first
and last cycle, respectively. For high pressure, ($p>130$ Pa), the EMT model overestimates the confining pressure. During relaxation from $p_{max}$ down to about 50 kPa, the media is five times stiffer than during compression, where $S_{e}=580$ MPa. Under $p=50$ kPa the apparent elasticity decreases quickly, and tends to zero.

EMT gives an expression for the oedometric effective stiffness
$S_{e}=b^{3/2}\frac{\phi Ck_{n}}{6\pi}$ defined in eq. (\ref{eq: EMT pres dep}), with $b=9/10$ for a friction-less and $b=43/30$ when friction is included (see Appendix \ref{sec:Apendice-B-EMT}). The coordination number,
or the average number of contacts per particle, is estimated to be
$C=6$ in a friction-less 3D packing \citep{makse1999effective}. For glass beads,
the normal effective stiffness is $k_{n}=145$ GPa; the oedometric effective stiffness should be $S_{e}=26$ GPa for a friction-less media, and  $S_{e}=52$ GPa when friction is included. The EMT predicts a stiffness two orders of magnitudes more than the experimental results. This gap between observations and a theory based on granular mechanics is also observed in soils \citep{diDonna2015response}; and suggests that, in the low-pressure range ($p<200$ kPa), cohesion and plastic deformation play an important role in the quasi-static response of a granular media. Adding friction to the EMT model gives a stiffness two times higher than a friction-less media. This suggests that friction is not the only reason why elasticity is five times higher during relaxation than compression.

The redirection factor can be estimated from the effective modulus ratio according to different hypothesis (see Appendix \ref{sec:Apendice-B-EMT}). In the case of a friction-less
media ($k_{s}=0$), consisting of perfectly smooth spheres, the coefficient of redirection is  calculated as $K=1/3$. The opposite limit for infinite friction between grains corresponds to $K=0.02$.
The experimental estimation of 0.2 indicates that compression friction plays a limited role during compression. During relaxation, $<\sigma_{yy}>$
is constant from 150 kPa to 50 kPa, meaning that $K$ tends to zero.
This indicates that friction forces cannot be neglected during relaxation.

Explaining hysteresis requires reconsidering the hypothesis of no-slip ($\xi_{s}=0$)
or perfect slip ($k_{s}=0$) in eq. (\ref{eq:InterGrainForce}). To
do so, a kinematic friction coefficient $\mu$ and viscosity $\gamma$
between the spheres should be introduced. The sliding condition requires
computing $\xi_{s}$ for every time step to check for sliding. In
this case the inter-grain force can be written as a function of the
normal and tangential components of the displacement \citep{makse2004granular}:
\begin{equation}
\overrightarrow{F}=\left(\xi_{n}\overrightarrow{n}+\gamma_{n}\dot{\xi_{n}}\right)k_{n}\sqrt{\xi_{n}R_{eff}}\overrightarrow{n}+\textrm{min}\left(\mu F_{n},k_{s}\sqrt{\xi R_{eff}}\xi_{s}\right)\overrightarrow{s}.\label{eq: MD}
\end{equation}
\citep{garcia2006} compute numerical response of oedometric cycles with stress-strain
hysteretical curves for both axial and lateral stress qualitatively in agreement with Figs. \ref{fig:ssRaw} and \ref{fig:Compressive-stresses-averaged}.
They also retrieve the  evolution of the apparent elastic modulus $M$ due to compaction.  

 EMT model better explains the stress-strain curve during compression at low pressure up to $p\sim130$ kPa. Nevertheless, the absolute value of the oedometric effective stiffness is not in agreement with the experiment. The redirection factor indicates that friction is strong during relaxation, and weak during compression. The compaction changes the effective stiffness between each compression, but the relaxation path is very stable. Relaxation data, hysteresis and evolution of the response between each cycle can be explained quantitatively by numerical simulations with an inter-grain force expressed in eq. (\ref{eq: MD}).

In section \ref{subsec:Discussion} we will come back to the analysis of the quasi-static characterization compared to the dynamic one, which is presented in the next section.

\section{Dynamic characterization}\label{sec:Dynamic-characterization}


\begin{figure}
\centering \includegraphics[width=0.95\columnwidth]{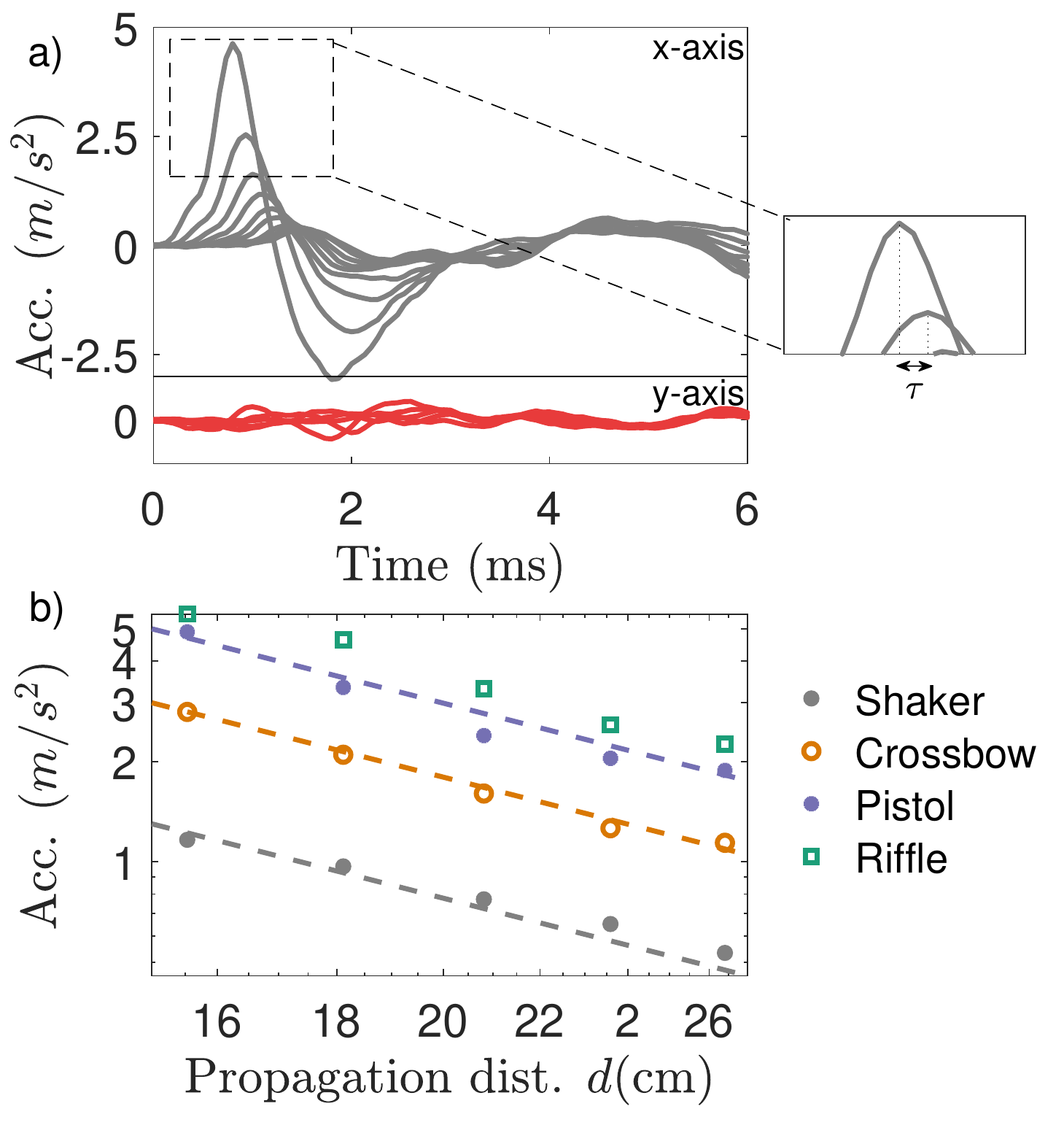}

\caption{\label{fig:TimeaSignals}a) Experimental time signals of a seismic wave produced by the impact of a projectile in a granular media. Ten accelerometers are distributed along the $\protect\overrightarrow{x}$-axis (from $x=$13 to 38 cm depth, see Fig. \ref{fig:scheme}). This example corresponds to a confining pressure of 20 kPa. At this low pressure,
the $\protect\overrightarrow{x}$-component of the acceleration (gray)
is one order of magnitude higher than the $\protect\overrightarrow{y}$-component (light red). The inset shows the time delay $\tau$ between two signals computed by cross-correlation to estimate the wave speed. b) Amplitude of the peak acceleration as a function of the propagation distance.
The peak acceleration, averaged over the confining pressure, is represented for the  different sources as indicated in the colored legend. The dashed lines correspond
to the numerical simulation, with an attenuation coefficient $\alpha=2.3$ Np/m (see Section \ref{subsec:Wave-propagation-numerical}).}
\end{figure}

Seismic waves traveling through granular media can be characterized
by recording the acceleration of the material (Fig. \ref{fig:TimeaSignals}a). The $x$-component is the main component of the wave propagating along the $\overrightarrow{x}$-axis, but a $\overrightarrow{y}$-component is also recorded. The direct wave propagation is observable from 0 to 2 ms. 

An average of the maximum amplitudes over the different confining pressures is presented in Fig. \ref{fig:TimeaSignals}b. The propagation distance is defined as:  ${d=\sqrt{x{^{2}}+y_{0}^{2}}}$. 
An exponential decrease is observed with amplitudes proportional to $d^{-n}$, with $n=2.2$ for the pistol and $n=2.4$ for the crossbow. In the case of Riffle, a sensor saturation occurs above ${6\textrm{ m/s}^{2}}$. We observe a strong wave attenuation compared to the case of an infinite non-attenuated media, for which $n=1$, corresponding to the conservation of energy along a spherical surface.

The actual experiment has three main difference compared the fore-mentioned ideal case. First, the media has a finite size of around a wavelength, where interference between incident and refracted waves occurs. Second, in this sub-wavelength region P- and S-waves interact together in a near field term with a $n=2$ slope \citep{Aki_1980}. Third, the amplitude decrease is also driven by the source extension that cannot be considered punctual compared to both the size of the media and the wavelength. The complexity of propagation in this sub-wavelength region is observed experimentally in the following Section \ref{subsec:Pression-dependant-wave-properti} and numerically modeled in section  \ref{subsec:Wave-propagation-numerical}.

\subsection{Pressure-dependent wave properties\label{subsec:Pression-dependant-wave-properti}}

\begin{figure}
\centering \includegraphics[width=0.95\columnwidth]{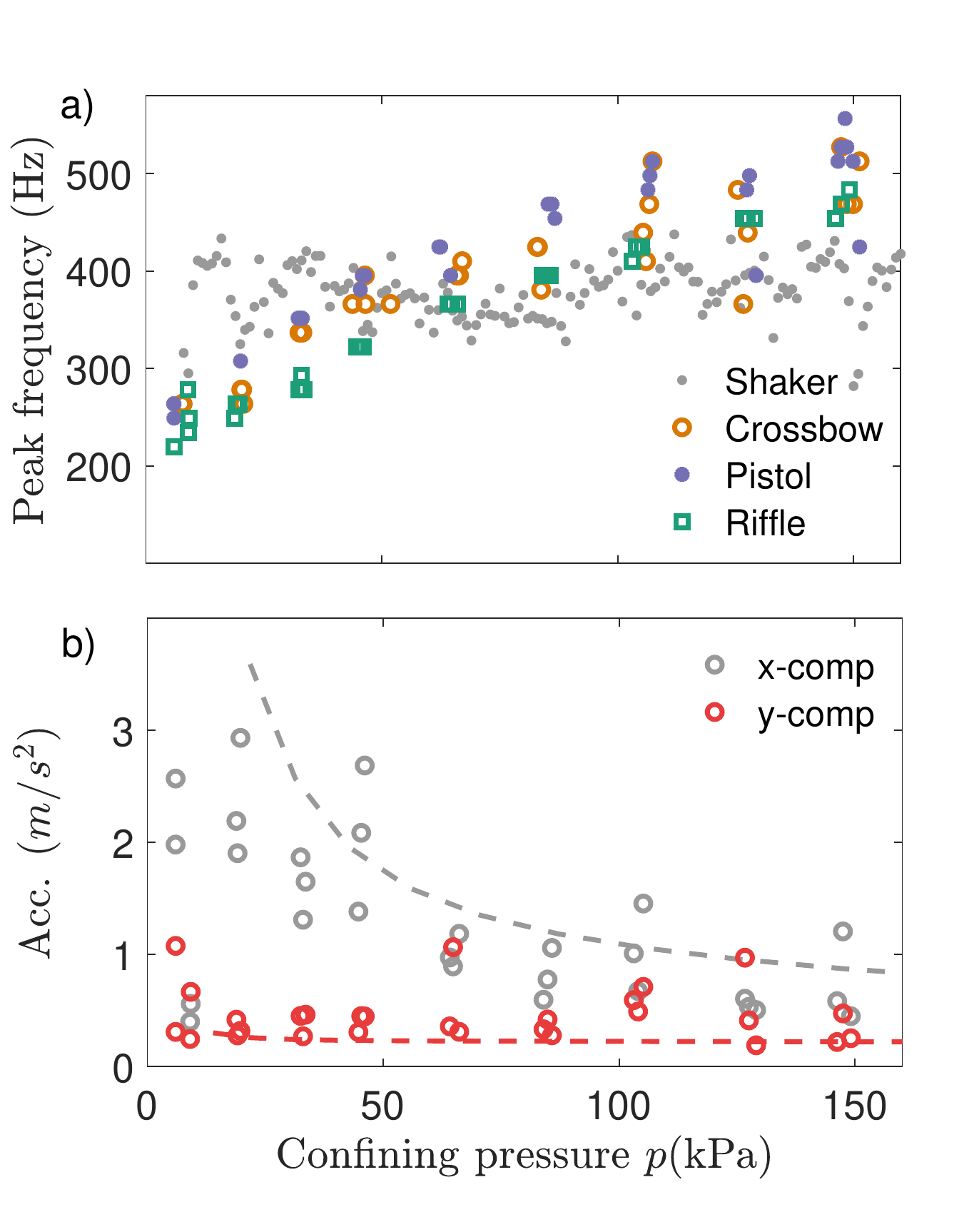}

\caption{\label{fig:WavePropPres} \textbf{a)} Peak frequency as a function
of confining pressure for glass beads. Different symbols represent experiments with different projectiles, guns, and the shaker. 
\textbf{b)} Maximum acceleration as a function of confining
pressure for crossbow signals. The $\protect\overrightarrow{x}$-component of the maximum acceleration is shown in orange, and the $\protect\overrightarrow{y}$-component in magenta. The dashed lines represent acceleration peaks for simulated waves (see Section \ref{subsec:Wave-propagation-numerical}).}
\end{figure}

The shaker input signal is a Heaviside step function resulting in
a wave propagating, with an average frequency peak of $\sim400$ Hz,
in the whole pressure range from 7 to 160 kPa (see Fig. \ref{fig:WavePropPres}a).
Impact-born waves present a center frequency ranging from 200 Hz to 500 Hz. It appears that the projectile energy (see Table \ref{table:guns}) doesn't change significantly the frequency content after ten centimeters of propagation.

In Fig. \ref{fig:WavePropPres}b, peaks acceleration of the $\overrightarrow{x}$ (gray dots) and $\overrightarrow{y}$ (red dots) components at $x=16$ cm, are represented for different confining pressures. The $\overrightarrow{x}$-component of the maximum acceleration is shown to decrease with pressure, while
the $\overrightarrow{y}$-component is much smaller and nearly constant. Experimental amplitudes are compared to simulated data (dashed lines
in Fig. \ref{fig:WavePropPres}b, with the same color code for the $\overrightarrow{x}$ and $\overrightarrow{y}$ components) (see section \ref{subsec:Wave-propagation-numerical}).

\begin{figure}
\centering \includegraphics[width=0.95\columnwidth]{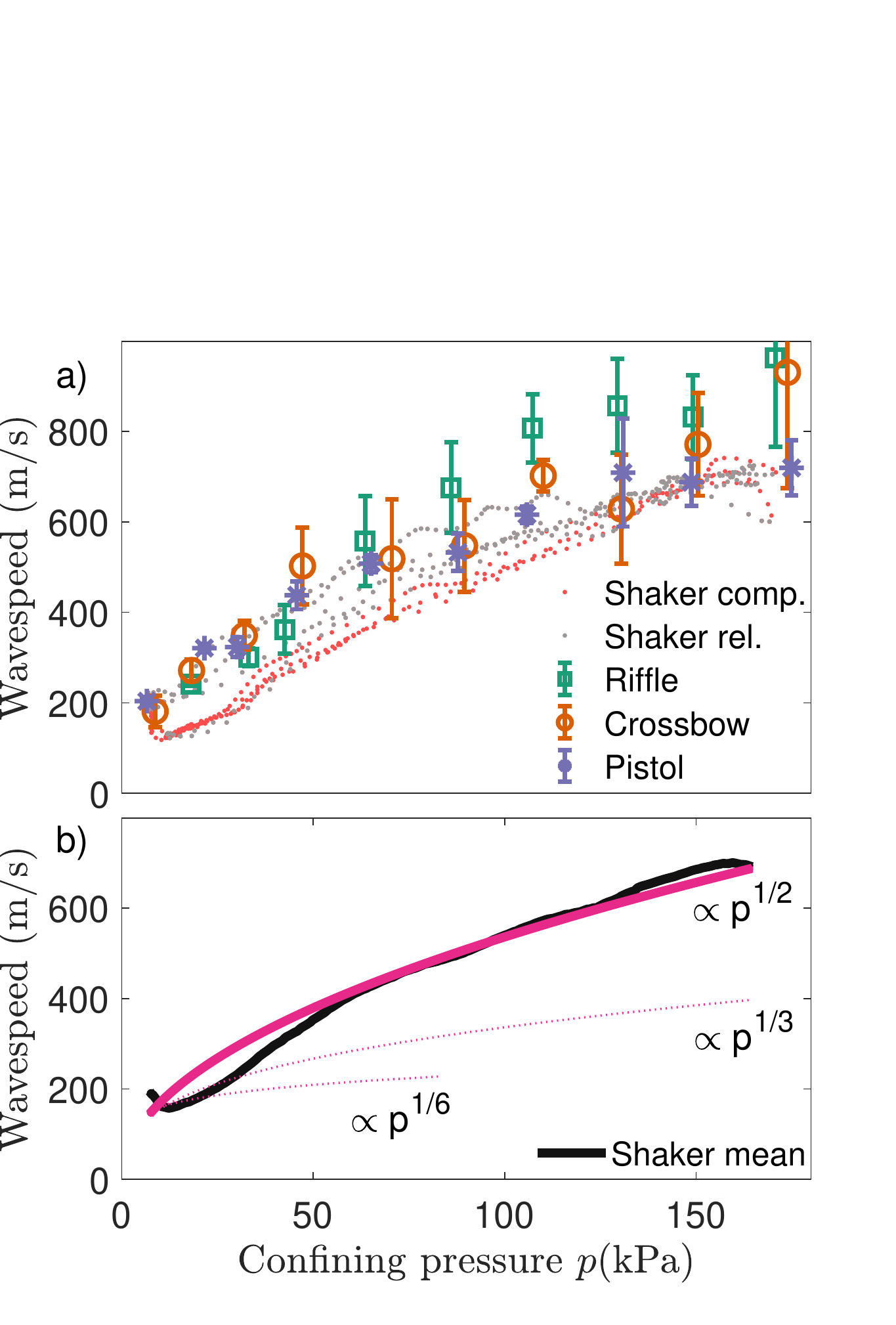}
\caption{\label{fig:Wavespeeds} a) Wave speed as a function of the
confining pressure. Different symbols represent experiments with different projectiles, guns, and the shaker. For the shaker, we show data during several runs of the compression (gray dots) and relaxation (red dots) phases (see the cycles in Fig. \ref{StaticMeasurement}). For the different impacts, we compute the mean and standard deviation in each pressure step. b) The average shaker velocity as a function of the confining pressure is shown with a thick black line. 
Trends lines with $p^{1/6}$,  $p^{1/3}$ and  $p^{1/2}$ dependencies are shown in fuchsia corresponding to EMT models in dashed lines and mesoscopic model with $\beta_{c}=4.7\times10^{3}$ (or equivalently modified EMT model) in full line. }
\end{figure}

The wave speed ($V=\delta x/\tau$) was measured by estimating the time lag ($\tau$) between two sensors signals separated by $\delta x$. An estimate of the time lag $\tau$ is shown in the inset of Fig. \ref{fig:TimeaSignals}. The maximum of the cross-correlation is estimated with quadratic interpolation to improve precision. This method is known to be very robust \citep{cespedes1995methods}.
All possible pairs of the five accelerometers within the measuring depth (15-27 cm) are used to estimate experimentally the wave speed using the cross-correlation method illustrated in Fig. \ref{fig:TimeaSignals}. The wave speed as a function of the confining pressure is shown in Fig. \ref{fig:Wavespeeds}a. For impacts, each wave speed is an average over five attempts at a constant pressure. Error bars represent the standard deviation.

A clear increase in wave speed from 200 m/s at $p_{min}$,
up to 800 m/s ($\pm 100$) at 160 kPa is observed in Fig. \ref{fig:Wavespeeds}a.
Despite the wide range in kinetic energy of the projectiles (see Table \ref{table:guns}), the wave speed measurements do not reflect a clear dependency on it.

Shaker-born waves present a very similar confining pressure
dependency than impact-born waves (see Fig. \ref{fig:Wavespeeds}a). Wavespeeds measured during compression are represented
with red dots, and those during relaxation with gray dots.
There is no measurable difference between compression and relaxation,
except at low pressure ($p<50$ kPa). Contrary to the stress-strain relationship in the quasi-static regime, there is no significant hysteresis in the wave speed measurements.

The apparent wave speed $V$ measured in Fig. \ref{fig:WavePropPres} extends from 200 to 800 m/s. The corresponding frequencies $f$ spans from 200 to 500 Hz according to Fig. \ref{fig:Wavespeeds}, and the wavelengths $\lambda=V/f$ ranges from 0.5 m to almost 2 m. These values are three orders of magnitude larger than the grains ($d\sim0.5$ mm, see Fig. \ref{fig:Grain size}). In this $\lambda/d>>10$ regime, coherent waves propagate in an equivalent homogeneous material  \citep{leGonidec2007multiscale}.
There is no  reflection at each grain contact and thus no multiple-scattering. A wavefront cannot sense an individual grain but an equivalent media. This is different from the diffusive regime of propagation where the energy spread and decay are related to a mean free path. This regime requires wavelength of the same order as the grain size \citep{langlois2015}.

\subsection{Wave propagation: numerical simulation \label{subsec:Wave-propagation-numerical}}

\begin{figure}
\centering \includegraphics[width=0.95\columnwidth]{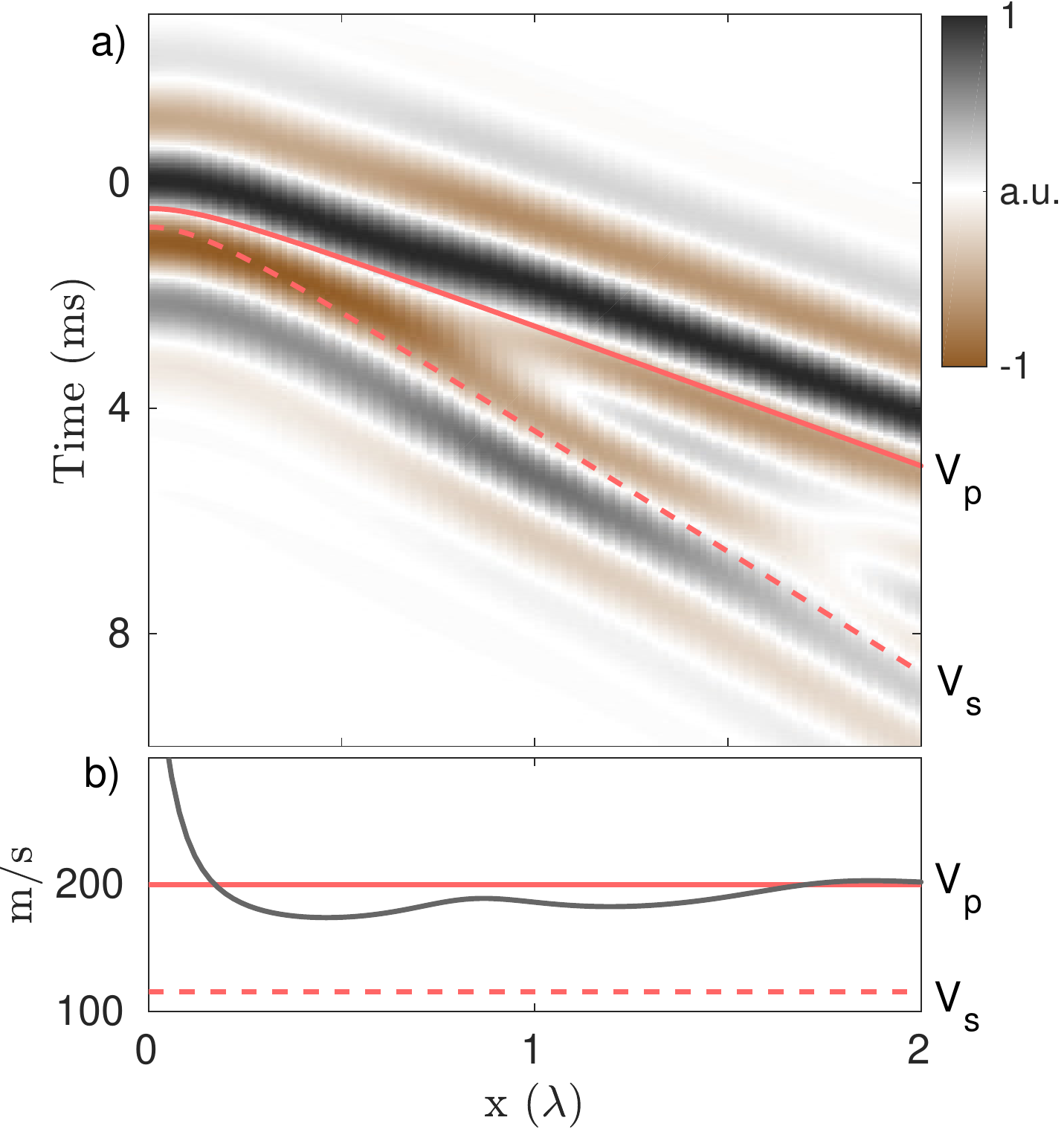}\caption{\label{fig:Time simul}a) Numerical simulation of an elastic wave
propagating in an infinite 3D media as a function of time and along
the $\vec{x}-$axis at $y=9$ cm from the source, with $V_{p}=200$
m/s, $V_{s}=115$ m/s, and the source frequency $f=400$ Hz. The color scale indicates the wave amplitude normalized at each depth (arbitrary units). $x$-coordinate
is normalised by the P-wavelength $\lambda=0.5$ m. The 8-cm diameter
source is centered at the origin. The propagation time is indicated
by a red full line for $V_{p}$ and a dotted line for $V_{s}$. b)
Black line represents the wave speed estimation based on the correlation
method. $V_{p}$ and $V_{s}$ are indicated in red full and dotted
lines respectively. In a near-field region ($x<\lambda/4$) the apparent
wave speed is higher than $V_{p}$ due to the extended source size.
In the far-field region ($x>\lambda/4$), P and S-waves begin to separate
and the estimated wave speed tends to $V_{p}$. Between these two
regions, a single wave is traveling at a wave speed smaller than $V_{p}$. }
\end{figure}

\begin{figure}
\centering \includegraphics[width=0.95\columnwidth]{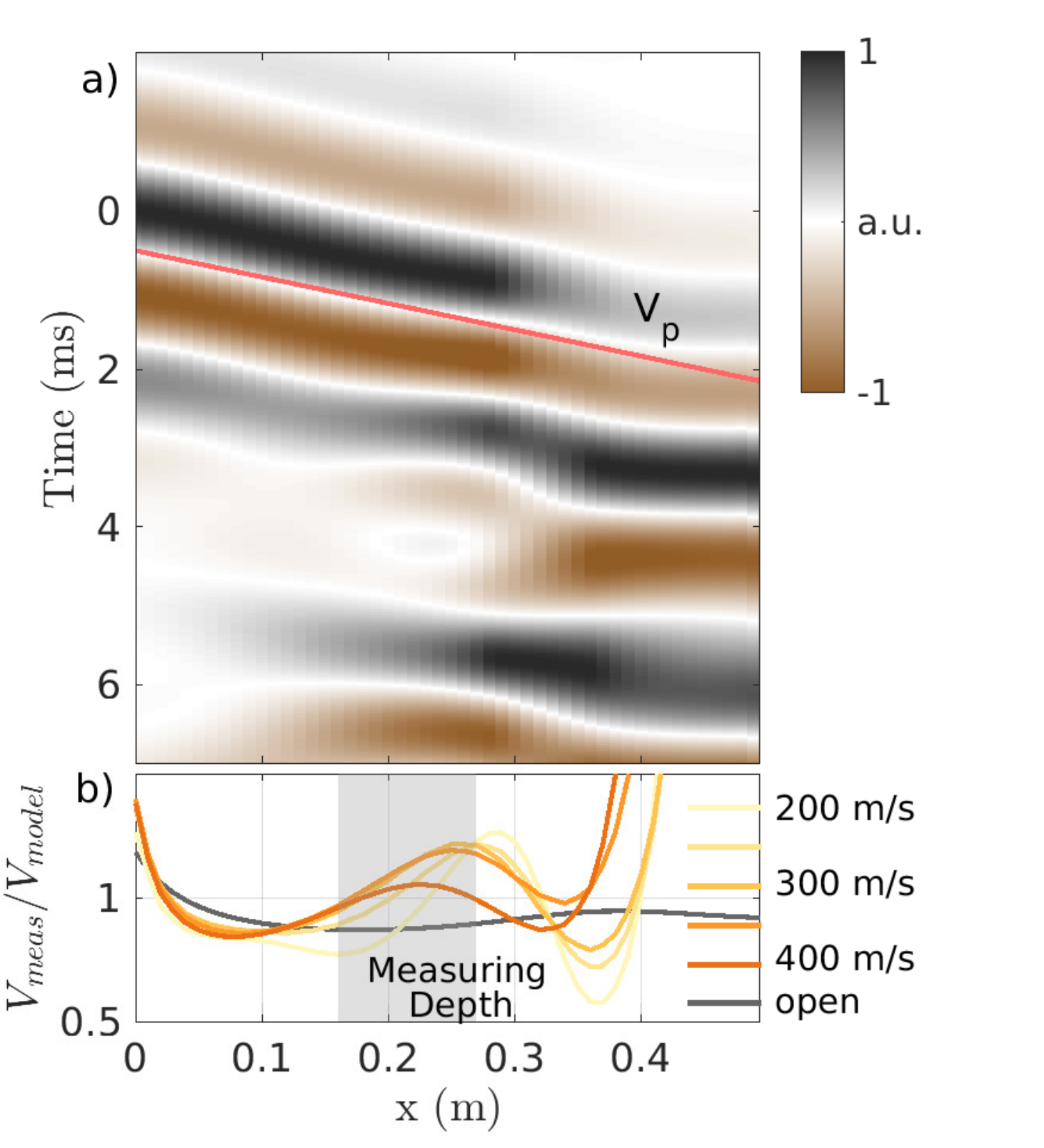}\caption{\label{fig:Closesimul}a) Numerical simulation of an elastic wave
propagating in a closed 3D media as a function of time and along the
$\vec{x}-$axis under the source ($y=0$ cm). The color scale indicates the wave amplitude normalized at each depth (arbitrary units). The red line indicates the propagation
time at the P-wave speed model $V_{model}=300$m/s. b) Apparent wave
speeds estimated by correlation are normalized by the model P-wave
speed ranging from 200 to 400 m/s as indicated with the color legend.
The wave speed in open media is represented in black for comparison
($\lambda=0.5$ m). The measuring depth gray zone indicates the accelerometer
position in the experiment. In this region, the averaged wave speed
estimation ranges from -2\% to 10\% of the P-wave speed model.}
\end{figure}


Taking into account the propagation regime described in the previous section, a conventional wave propagation simulation in a linear homogeneous elastic media was chosen. The granular nature of the media, important in the quasi-static characterization, is taken into account only when the confining pressure changes, not during the propagation phenomena. This hypothesis
implies that a simulation corresponds to a particular confining pressure. The wave propagation is considered to be linear, while the quasi-satic perturbation is nonlinear. This is a usual
hypothesis in the dynamic nonlinear characterization of rocks \citep{guyer2009nonlinear}.

Numerical simulations were performed using Matlab toolbox \textit{k-wave} \citep{treeby2018rapid,treeby2010k,treeby2012modeling}, based on a pseudo-spectral method. We use a 3D elastic code to fully understand the underlying wave physics. The medium is considered to be homogeneous with a density $\rho=1650\textrm{ kg/m}^{3}$. EMT gives a P- to S- ratio $V_p/V_s=\sqrt{3}$ for the friction-less case
(see Appendix \ref{sec:Apendice-B-EMT}). The P-wave speed is set
between 200 to ${500\text{ m/s}}$. Considering the Janssen model (see Appendix \ref{sec:Apendice-Janssen}), even at low external pressure a homogeneous media is suitable to understand the experiment.

The impact of the projectile is considered as a temporal force. Crater
properties are not studied in detail \citep[see \textit{e.g.}][]{crassous2007impact}.
Instead, we focus on body wave propagation. The force is applied along
the $\overrightarrow{x}$-axis on an 8cm diameter disc in the $\overrightarrow{z}$-$\overrightarrow{y}$
plane, centered at the coordinate's origin. The source radius was chosen as an estimate of the crater size. Between 4 and 10 cm, the source diameter does not influence the quantities of interest (relative amplitude and apparent wave speed). The source was set as a time dependent force: a Gaussian pulse with a 0.7 bandwidth centered at 400 Hz, according to Fig. \ref{fig:WavePropPres}. Sensors are distributed along the
$\overrightarrow{x}$-axis, and at a distance $y_{0}=9$ cm from the impact center. The time step is set automatically (form 5 $\textrm{\ensuremath{\mu}s}$
or 15 $\textrm{\ensuremath{\mu}s}$, depending on the frequency and wave speed), and the spacial grid step is set at 1 cm.

Elastic propagation in a 3D open homogeneous media is handled thanks
to a large Perfectly Matched Layer (20 points). The $\overrightarrow{x}$-component
of the particle velocity, $u_{x}\left(x,t\right)$, as a function
of time and depth, is presented in Fig. \ref{fig:Time simul}a, for a P-wave speed of $V_p=200$ m/s. 
To help visualization, the waveform is normalized by $A\left(x\right)$, the  maximum amplitude at each depth. Fig. \ref{fig:Time simul}b
represents the estimated wave speed by cross-correlation as a function
of depth. Close to the source position, the wave speed is above $V_{p}$,
due to the extended source size. In the sub-wavelength region ($x<\lambda$),
the apparent wave speed is only 87\% of $V_{p}$, because of the interaction
with the S-wave. This 13\% difference depends on the sensor position along the $\overrightarrow{y}-$axis,
since the compressive-to-shear amplitude ratio varies according to the relative source position. At $x=\lambda$, pressure and shear waves begin to separate and the apparent wave speed increases and reaches $V_{p}$ around $2\lambda$. 

The confining pressure dependency can be introduced in the simulation code by changing the wave speed according to the measurements represented in Fig. \ref{fig:Wavespeeds}b. We add to Fig \ref{fig:WavePropPres}a the simulated maximum amplitudes in dotted lines. The same arbitrary amplitude is set for each wave-speed (or pressure) simulation. Numerical simulations show a similar trend for both components, which is explained only by the wavelength growth and the relative P- to S-wave contribution at this particular spot.

The maximum amplitudes computed numerically at each depth are shown in Fig. \ref{fig:TimeaSignals},
including an attenuation in the form ${A_{\alpha}\left(d\right)=e^{-\alpha d}A\left(d\right)}$,
with $\alpha=2.3$ Np/m. This attenuation coefficient is 4 orders of magnitude higher than in
rocks \citep{liu2020elastic}. In similar conditions, attenuation measurement
is $0.6\pm0.6$ Np/m for S-waves at 450 Hz in water-saturated sand
\citep{brunson1980laboratory}, and 0.15 Np/m in dry sand at 500 Hz
\citep{koerner1976acoustic,leinov2015investigation}.

The effect of a closed media on an elastic wave with no attenuation
was studied. Abrupt changes in the mechanical properties are not easily
handled from a numerical point of view. The normal incidence reflection
of P-waves is characterized by the contrast impedance ($Z=\rho V_{p}$)
between two media ($r_{12}=Z_{1}/Z_{2}$). Densities are similar for
the granular media and the acrylic walls. The wave speed in acrylic is $\sim1000$ m/s. The contrast impedance between acrylic and granular media ranges from 0.2 to 0.7. The 14-mm thick acrylic walls (two orders
of magnitude smaller than $\lambda$) have almost no effect on the propagation; including the walls in the simulation results in a negligible 2\% perturbation of the wave speed. On the other hand, air is three orders of magnitude lighter than the granular media, meaning an impedance ratio of $4\times10^{2}$. Such a contrast
cannot be handled numerically. Instead, the surrounding media has both density and wave speed ten times smaller than the propagating media achieving a $10^{2}$ impedance contrast. This corresponds to a reflection of 2\%  of the wave amplitude instead of 1\% with air.

In such a closed media the relative amplitude decay in the region of interest is the same as in an open space. An interference between
$\overrightarrow{x}$-negative and $\overrightarrow{x}$-positive
propagation direction is seen in Fig. \ref{fig:Closesimul}a. Fig.
\ref{fig:Closesimul}b represents the apparent wave speed as a function
of depth estimated by cross-correlation, for a variety of modeled wave speeds. The estimated wave
speed from the numerical simulations are normalized by the P-wave speed
value. The positions of the accelerometers in the experimental set-up are indicated with a gray zone. At this depth range, the averaged wave speed shows a difference of 3\%, 9\%, and -2\% between the measured wave speed and the model P-wave speed, being 200, 300 and 400 m/s respectively .

The numerical analysis demonstrates that despite the interferences in a closed elastic medium, we measure the actual P-wave
speed in the experimental configuration with an error smaller than
10\%. In addition, there is no trend with the absolute P-wave speed of the model. In conclusion,
this numerical study ensures confidence in the experimental estimation of the mechanical parameters of the granular media. The following section focuses on the analyses of the  P-wave speed dependency on the confining pressure represented in Fig. \ref{fig:Wavespeeds}b.

\subsection{Discussion on the elasticity pressure dependency properties\label{subsec:Discussion}}

Most of the literature reports a $p^{e}$ dependency on the wave speed
(P or S-wave), with an exponent $e$ between $1/4$ to $1/6$ \citep{zimmer2007seismic,garcia2006,jia2021elastic},
as it stands in the EMT P-wave speed expression \citep{makse1999effective}:
\begin{equation}
V_{emt}=\frac{3}{\sqrt{10\rho}}\left(\frac{\phi k_{n} C}{6\pi}\right)^{1/3}p^{1/6},\label{eq:EMT}
\end{equation}
where $C$ is the coordination number or the average number of contacts
per particle. 
Taking into account other effects, the EMT trend reaches
a ${1/3}$ exponent \citep{goddard1990nonlinear,wichtmann2004influence,agnolin2007internal,zimmer2007seismic}. These
trends are represented in Fig. \ref{fig:Wavespeeds}b, where $C$
is computed to fit the data at $p_{min}.$ Any pressure value can
be chosen to compute $C$. In Fig. \ref{fig:Wavespeeds}b we present several curves trend curves with different exponents (1/2, 1/3, 1/6), all of them fitted at $=p_{min}$. We observe that $e=1/2$ is the best fit. $e$ between 1/3 and 1/6 show considerable departures respect to the measurements.

We propose to modify the EMT of eq. (\ref{eq:EMT}) by including a pressure
dependency on the coordination number as $C=\frac{p}{p_{max}}C_{max}$.
The resulting wave speed is:
\begin{equation}
V_{mod}=\frac{3}{\sqrt{10\rho}}\left(\frac{\phi  k_{n} C_{max}}{6\pi p_{max}}\right)^{1/3}p^{1/2}.\label{eq:EMTMod}
\end{equation}
The modified EMT suits very well the observations, as shown in Fig. \ref{fig:Wavespeeds}b
in fuchsia. The coordination number is changing from 0.6 to 13. This
variation seems exaggerated but it is in agreement with experimental
measurements in unconsolidated sand with variations from 2 to 18 \citep{wright2021coordination}.
Nevertheless, such an increase of contacts per grain is explained by the authors
by a filling factor changing from 0.3 to 0.8, while in the present
measurement the variation $\mathrm{d}\phi=\epsilon\sim3\%$ is negligible.
The lowest 0.6 contact number is not physically possible. Instead of
this modified EMT, we propose another approach to explain the data.


Rocks are an aggregate of minerals and can be considered cemented
granular media. Elasticity in rocks presents mesoscopic nonlinearity based on nonlinear elastic energy considerations. The elastic energy can be written as function of three invariant of the  Lagrangian strain.  At the third order in energy, the elasticity of a solid is $M_{solid}=M_{o}\left(1+\beta_{c}\epsilon\right)$,
with $\beta_{c}$ the third order nonlinear parameter defined with a positive compressive strain. $M_{0}$ is the
elasticity with no external perturbations \citep{guyer2009nonlinear}. Granular media, as contrary to rocks, do not have any elasticity without external force. A mesoscopic nonlinearity for granular media is then $M=M_{o}\beta_{c}\epsilon$. Including a non-hysteretical linear pressure-strain
relationship, $\epsilon=p/M_{0}$ gives a simple dependency of $M$ with pressure:
\begin{equation}
M=\beta_{c}p. \label{Eq:NLElastGrain}
\end{equation}
The associated wavespeed $V^2_{meso}=M/\rho$  is then deduced: 
\begin{equation}
V_{meso}=\left(\frac{\beta_{c}}{\rho}p\right)^{1/2}.\label{eq:wavespeed mesoscopic}
\end{equation}
This expression gives the same trend as the modified EMT. The $\beta_{c}$
parameters were computed by minimizing the sum of the squared difference
between the model and data. A reasonable agreement is observed between
experimental data and the wave speed computed with eq. (\ref{eq:wavespeed mesoscopic}),
with $\beta_{c}=4.7\times10^{3}$.

\begin{figure}
\centering \includegraphics[width=0.95\columnwidth]{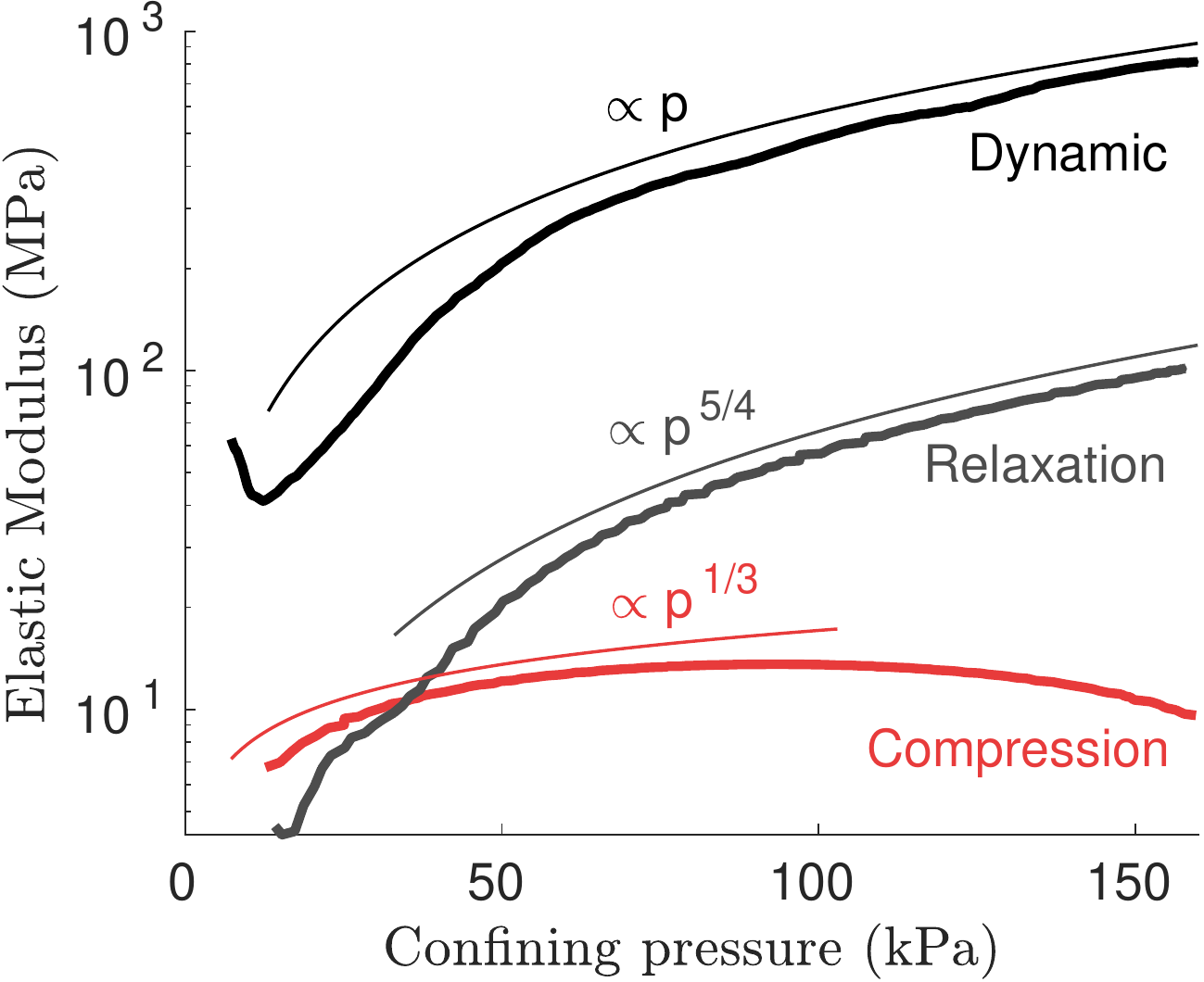}
\caption{\label{fig:Elasticity} Elastic modulus as a function of the confining pressure measured through wave propagation (Dynamic in black), quasi-static compression (red), and quasi-static relaxation (gray). Pressure dependency trends indicated in thin lines corresponds to respectively a mesoscopic nonlinearity or equivalently a modified EMT, EMT and empirical fit. }
\end{figure}

Experimental results from quasi-static (Fig. \ref{StaticMeasurement}) and dynamic acquisitions (Fig. \ref{fig:Wavespeeds}) can be represented in the same plot by computing the elastic modulus from the P-wave speed,
$M_{d}=\rho V_{p}^{2}$ and stress-strain curves  $M_{s}=\frac{\partial p}{\partial\epsilon}$.
Fig. \ref{fig:Elasticity} represents the dynamic and
the quasi-static elasticity for both compression and relaxation. Tendency
curves are also represented in thin lines. These are based on eq.
(\ref{Eq:NLElastGrain}) for the dynamic curve, and eq. (\ref{eq: EMT pres dep})
for compression. The $p^{5/4}$ dependency of the relaxation is purely
empirical. The elasticity difference between compression and relaxation,
already discussed in Section \ref{subsec:Data interpretation}, mainly
originates from friction differences and energy storage.

The dynamic elasticity is at least one order of magnitude above the quasi-static elasticity. We should quantify the mechanical perturbations of these phenomena to discuss this important difference. The times scales are 30 s for an oedometric cycle and 1ms for half a period for the dynamic wave. The strain scale is not straightforward and require to use the numerical simulation. The volumetric stress $\sigma_{v}=\sigma_{ii}/3$ (with  Einstein summation convention), can be set as an output of the numerical wave propagation model (see Section \ref{subsec:Wave-propagation-numerical}).
The compressive strain is then $\epsilon_{v}=\sigma_v/V^{2}_{p}\rho$. The strain is found to be $\epsilon_{v}=1.4\times10^{-4}$ in the vicinity of the source ($x=0$) for the higher strain scenario: riffle impact with $V_{P}=200$ m/s.  This dynamic stress should be compared to a volumetric strain computed from the quasi-static axial strain $\epsilon_{xx}$: $\epsilon_{v}=\overline{\epsilon_{xx}}\left(1+2K\right)/3$. With $3\%$ a maximum axial strain and $K=0.2$, the maximum volumetric strain is $\epsilon_{v}=1.4\times10^{-2}$.

In conclusion the  quasi-static perturbation
is ten thousand times slower and a hundred times larger in strain than the dynamic one. These order of magnitudes explains why no hysteresis is observed during wave propagation because sliding requires more stress and grain reorganization takes more time. These phenomena occur only during quasi-static compression. Furthermore, EMT that includes neither slipping nor reorganization predicts an elasticity in the same order of magnitude than the dynamic measurement ($M_{s}=380$ at $p=80$ kPa). This suggests that the dynamic measurement probes the grain contact mechanics, while quasi-static relates preferentially to the sliding and cohesion effect. 

\subsection{Results in natural media\label{sec:differentmedia}}

\begin{figure}
\centering \includegraphics[width=0.95\columnwidth]{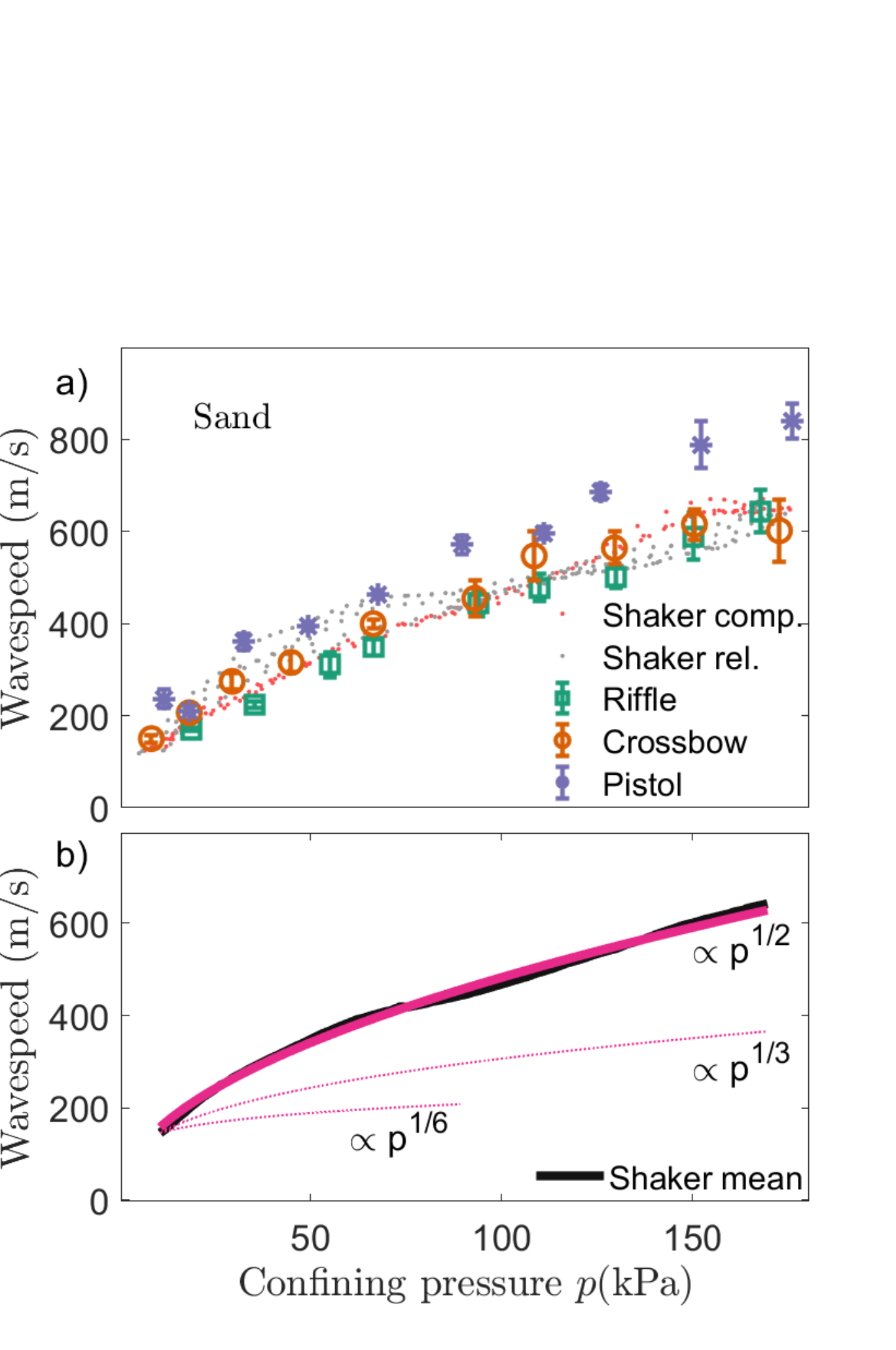}
\caption{The wave speed as a function of the confining pressure for sand. Similar caption as Fig. \ref{fig:Wavespeeds}. The the mesoscopic model has $\beta_{c}=3.9\times10^{3}$. \label{fig:WS_Sand}}
\end{figure}

\begin{figure}
\centering \includegraphics[width=0.95\columnwidth]{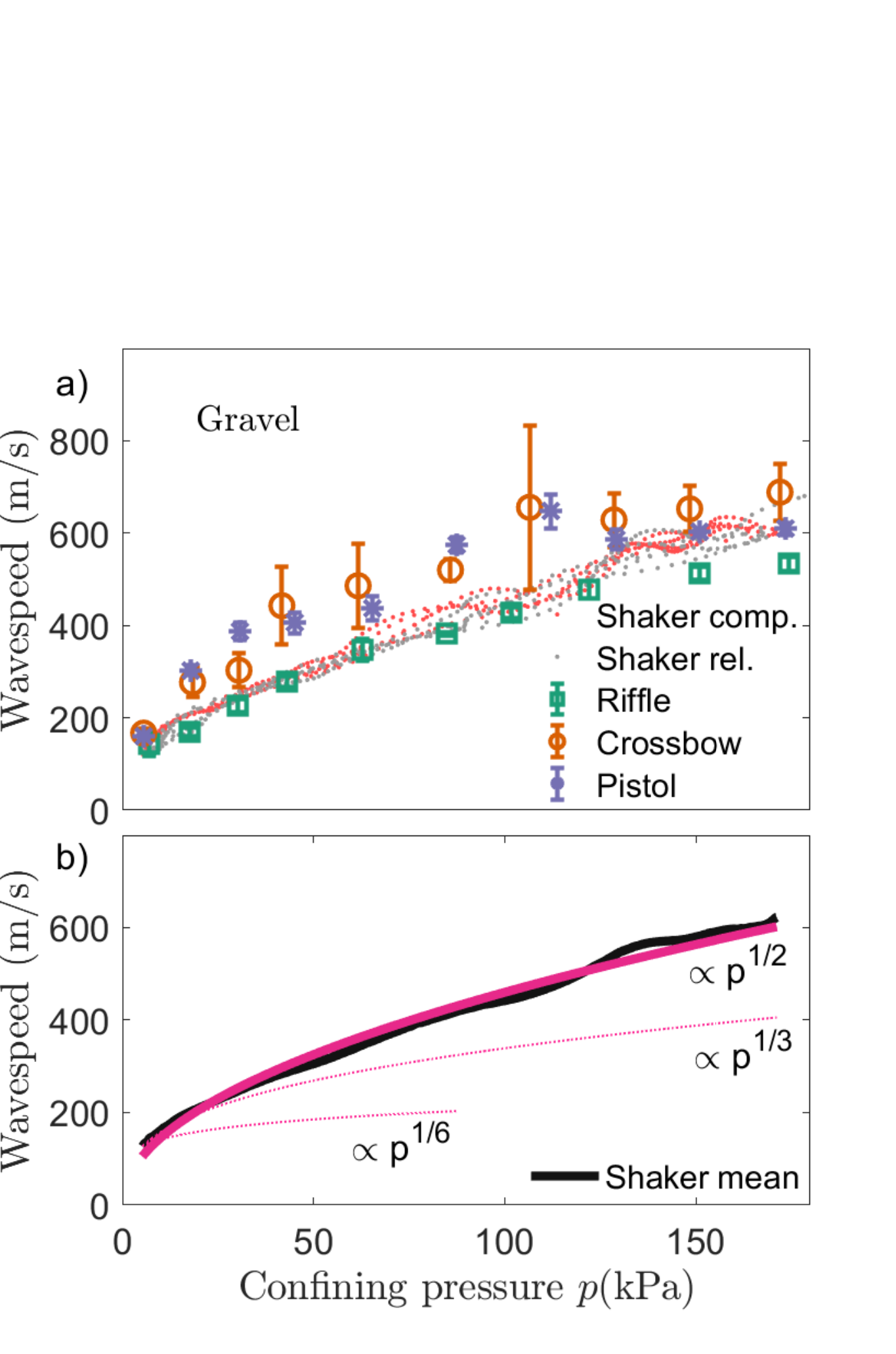}
\caption{The wave speed as a function of the confining pressure for gravel.  Similar caption as Fig. \ref{fig:Wavespeeds}. The the mesoscopic model has $\beta_{c}=3.5\times10^{3}$.}
\label{fig:WS_Gravel} 
\end{figure}


The wave speed as a function of confining pressure is shown in Figs. \ref{fig:WS_Sand}a and \ref{fig:WS_Gravel}a for sand and gravel, respectively. Data analysis is the same as for glass beads, explained in Section \ref{subsec:Pression-dependant-wave-properti} and there were no significant differences between the materials on the peak frequencies or amplitudes of Figs \ref{fig:TimeaSignals} and \ref{fig:WavePropPres}.
Trends are the same in natural granular media as in glass beads.
In Figs.\ref{fig:WS_Sand} b
and \ref{fig:WS_Gravel}b, we show the average of the wave speed for
the shaker with a black line, along with the mesoscopic model with a fuchsia line. We confirm the $p^{1/2}$ dependency of the wavespeed. The nonlinear $\beta_{c}$ parameters measured in sand
and gravel are $\beta_{c}=3.5\times10^{3}$ and $\beta_{c}=3.9\times10^{3}$,
respectively. These estimates have the same order of magnitude than measurements in
a dense granular media by a nonlinear harmonic generation with ultrasound \citep{brunet2008transitional}, with $\beta_{c}$ between 500 to 1500.
For consolidated granular material, the $\beta_{c}$ parameter is found to be an order of magnitude smaller: \textit{e.g.} for concrete is ($\beta_{c}\sim100-200$)
\citep{spalvier2020torsional}, and in rocks is between 0.5 to $2\times10^{3}$
\citep{gallot2015characterizing,DAngelo2008}. This indicates
that nonlinearity seems to decrease with grain's angularities.


\section{Conclusion}

We performed a mechanical characterization of confined granular media for both a quasi-static and dynamic (~500 Hz) regime.
During each quasi-static compression in the $50 - 200$ kPa confining pressure range, a quasi-elastic behavior with a pressure-constant elasticity was measured. EMT better explain the observations in the  $0 - 130$ kPa confining pressure range, with $p^{1/3}$ elasticity dependency. Each compression-relaxation cycles result in an increase of the elasticity. 
It was observed that the quasi-static response of the material can be well described by a quasi-elastic model \citep{Gennes1999}, exclusively during compression.  
The estimation of the redirection factor indicates that friction is only important during relaxation and can be neglected during compression. 
On the other way, relaxation reveals a highly hysteretic stress-strain relationship. Relaxation follows neither a quasi-elastic nor an EMT. We observe an elastic energy storage during the beginning of the relaxation; which is then released at lower pressure.

For each confining pressure step, impacts and vibrations in granular media generate highly attenuated elastic waves propagating in a homogeneous media with similar characteristics. The absence of a wave diffusion-like process is due to a difference of three orders of magnitude between grain size and wavelength.
The granular nature of the material is only relevant to understand the confining pressure incidence on the equivalent homogeneous elasticity of the media. This conclusion is supported by the agreement between numerical simulation models and experiments regarding wave speed, amplitude, and polarization of the elastic wave.
We also demonstrate that the wavefield is mainly compressional, propagating at the P-wave speed. The impact-generated waves are shown to be P-waves, with a dependency with the confining pressure of the type: $p^{1/2}$. We also present similar results of wave properties generated by impacts and shaker are very similar. Finally, the P-wave speed is shown to increase more than expected with a EMT model with a constant coordination number, $C$. A modified EMT model, with a $C$ proportional to the pressure, results in the same pressure dependency : $p^{1/2}$,  as a mesoscopic nonlinear model based on third-order nonlinear elastic energy. Both models fit the observations. The interpretation of these models is a deep reorganization in the particle contact network during the quasi-static perturbation.

\section*{Acknowledgements}


The authors acknowledge financial support from project FCE-1-2019-1-156451 of the Agencia Nacional de Investigaci\'on e Innovaci\'on ANII (Uruguay), the grant II-FVF-2019-145 of the "Fondo Vaz Ferreira" (D2C2-MEC), the project Grupos I+D Ciencias Planetarias C630-348 of the Comisi\'on Sectorial de Investigaci\'on Cient\'ifica (Udelar, Uruguay), and the Programa de las Ciencias B\'asicas (PEDECIBA-MEC, Uruguay).

\section*{Data Availability}
All our data are available upon reasonable request to the corresponding author. The k-wave package is available on http://www.k-wave.org/ under the GNU LGPL license.

\bibliographystyle{gji}
\bibliography{BiblioGranulaire}

\appendix
\section{Microscopic Quasi-elasticity} \label{sec:Apendicequasielasticity}

If we consider granular media as a homogeneous linear medium at the
macroscopic scale (average notation is omitted in this section), typically
larger than ten times the average size of a grain, the Hooke's law is: 
\begin{equation}
\sigma_{ij}=\lambda\delta_{ij}\epsilon_{ii}+2\mu\epsilon_{ii}.
\end{equation}

With an axi-symmetry around the $\overrightarrow{x}$-axis ($\sigma_{zz}=\sigma_{yy}$
and $\epsilon_{zz}=\epsilon_{yy}$), it becomes: 
\begin{equation}
\begin{cases}
\sigma_{xx}=\left(\lambda+2\mu\right)\epsilon_{xx}+2\lambda\epsilon_{yy}\\
\sigma_{yy}=2\left(\lambda+\mu\right)\epsilon_{yy}+\lambda\epsilon_{xx}
\end{cases}.\label{eq:3}
\end{equation}
Considering non-deformable walls along the $y$ and $z$-direction
($\epsilon_{yy}=\epsilon_{zz}=0$), the previous equation becomes:
\begin{equation}
\begin{cases}
\sigma_{xx}=\left(\lambda+2\mu\right)\epsilon_{xx}\\
\sigma_{yy}=\lambda\epsilon_{xx}
\end{cases},\label{eq:3-2}
\end{equation}

In this ideal case, the ratio between $\sigma_{xx}$ and $\epsilon_{xx}$
is the P-wave modulus: $M=\lambda+2\mu$, as written in eq. (\ref{eq:Elastic modulus}). Eq. \ref{eq:3-2} also gives an expression of the redirection factor (eq. \ref{eq:Poisson}):
\begin{equation}
K=\frac{\sigma_{yy}}{\sigma_{xx}}=\frac{\lambda}{\lambda+2\mu}. \label{eq:K}
\end{equation}
From this definition we can compute the Lame's coeficients in function of $M$ and $K$:
\begin{equation}
\begin{cases}
\lambda=KM\\
\mu=\frac{1-K}{2}M
\end{cases},\label{eq:lambamu}
\end{equation}
In order to evaluate the hypothesis of non-deformable walls, let's
consider a wall deformation as an equivalent spring along the wall:
$\sigma_{yy}=S\epsilon_{yy}$, with $S$ being the equivalent stiffness.
The redirection factor $K_{S}$, within the granular media, is then:
\begin{equation}
K_{S}=K\frac{1}{1-\frac{E}{S}\left(1+K\right)},\label{eq:RedeirS}
\end{equation}
with $\sigma_{yy}=K_{S}\sigma_{xx}$. If the spring stiffness is much
stronger than the Young modulus $E$, the wall can be considered as
non-deformable ($K_{S}\approx K$). From eqs (\ref{eq:sigmayy}) and (\ref{eq:Poisson})
redirection can be estimated as $K_{s}\approx0.2$ in glass beads,
and we know that the strain is lower with a wall deformation: $K<K_{S}$.
A wall deformation also redirects the strain along the $\overrightarrow{y}$-axis:
\begin{equation}
\epsilon_{yy}=-K_{\epsilon}\epsilon_{xx},\label{eq:8-1-1}
\end{equation}
with $K_{\epsilon}=\frac{\lambda}{\left[2\left(\lambda+\mu\right)-S\right]}$.
This should create a displacement along the $\overrightarrow{y}$-axis,
but it is not detectable with our measurement system, with an uncertainty
order of $0.01\%$ in strain. Considering a maximum strain $\epsilon_{xx}$
of $1\%$, the order of magnitude of $K_{\epsilon}$ is $10^{-2}$.
The stress and strain ratio $M_{s}$, in the case of deformable walls,
is: 
\begin{equation}
M_{S}=M\left(1-2KK_{\epsilon}\right).\label{eq:elasticityDef}
\end{equation}
The wall deformation tends to underestimate the elastic modulus. Nevertheless,
the maximum variation is around $0.14\%$. We conclude that the non-deformable
wall hypothesis is reasonable.

\section{Effective medium theory}\label{sec:Apendice-B-EMT}

The Effective Medium Theory (EMT) considers that the macroscopic stress
is equal to the average stress on each grain \citep{makse1999effective}.
Effective quantities can be established such as the normal effective
stiffness:
\begin{equation}
k_{n}=\frac{4\mu_{g}}{1-\nu_{g}},\label{eq:eqstif}
\end{equation}
with $\mu_{g}$ and $\nu_{g}$ the shear and Poisson's ratio of the
grain material, respectively. Typical values of these parameters for glass are $\mu_{g}=29$
GPa and $\nu_{g}=0.2$. In the case of a friction-less media, the bulk
and shear moduli are: 
\begin{equation}
\begin{array}{c}
K_{e}=\frac{1}{2}\left(\frac{\phi Ck_{n}}{6\pi}\right)^{2/3}p^{1/3}\\
\mu_{e}=\frac{3}{10}\left(\frac{\phi Ck_{n}}{6\pi}\right)^{2/3}p^{1/3}
\end{array},\label{eqEMT moduli}
\end{equation}
with $\phi$ the volume fraction, and $C$ the coordination number
(the average number of contact per particle). 
This gives an effective moduli ratio $r_{e}=K_{e}/\mu_{e}=20/12$. 
 Recording the equation relating effective Lame's coefficients ($\lambda_e$, $\mu_e$) and $K_e$: $\lambda_e=K_{e}-\frac{2}{3}\mu_{e}$, and substituting in Eq. \ref{eq:K} gives a redirection factor that only depends on $r_e$: \begin{equation}
    K=\frac{\frac{3}{2}r_{e}-1}{\frac{3}{2}r_{e}+2}.
 \end{equation}  
This is  $K=1/3$ for friction-less granular media. Including tangential forces modify the shear modulus as:
\begin{equation}
\mu_{e}=\frac{k_{n}+\frac{3}{2}k_{s}}{20}\left(\frac{6\phi^{2}C^{2}}{k_{n}\pi^{2}}p\right)^{1/3}, \textrm{with } k_{s}=\frac{8\mu_{g}}{2-\nu_{g}}.
\end{equation}
In this case the moduli ratio becomes $r_{e}=\frac{K_{e}}{\mu_{e}}=\frac{5\left(2-\nu_{g}\right)}{3\left(5-4\nu_{g}\right)}=5/7$, 
and $K=1/43\approx0.02$.

The elastic modulus can be expressed as $M_{e}=K_{e}+\frac{4}{3}\mu_{e}=K_{e}\left(1+\frac{4}{3r_{e}}\right)$,
giving $M_{e}=\frac{9}{5}K_{e}$ for the friction-less EMT media, and
$M_{e}=\frac{43}{15}K_{e}$ when friction is included. During relaxation,
the apparent elastic modulus is varying, but no better agreement can
be found with the EMT. The effective elasticity $M_{e}$ can also
be used to established an effective stress-strain relationship $\overline{\sigma_{xx}}=p=M_{e}\overline{\epsilon_{xx}}$
with the following pressure dependence strain for quasi-static comparison:
\begin{equation}
\overline{\epsilon_{xx}}=\left(\frac{p}{S_e}\right)^{2/3},
\end{equation}
with the oedometric effective stiffness given by $S_e=\left(M_{e}^{3}/p\right)^{1/2}$ . The friction-less case gives $S_{e}=\left(\frac{9}{10}\right)^{3/2}\frac{\phi Ck_{n}}{6\pi}$,
while $S_{e}=\left(\frac{43}{30}\right)^{3/2}\frac{\phi Ck_{n}}{6\pi}$ when friction is included.
P- and S-wavespeed are given by $V_{P}=\sqrt{\frac{M_{e}}{\rho}}$
and $V_{S}=\sqrt{\frac{\mu_{e}}{\rho}}$ respectively. The wavespeed
ratio is then: 
\begin{equation}
\frac{V_{P}}{V_{S}}=\sqrt{M_{e}/\mu_{e}}=\sqrt{r_{e}+4/3}.
\end{equation}
For friction-less ratio $V_{p}/V_{s}=\sqrt{3}$, and $V_{p}/V_{s}=\sqrt{2}$
when friction is included. 

\section{Janssen model } \label{sec:Apendice-Janssen}

According to the Janssen model \citep{Ovarlez2003,Ovarlez2005}, the
apparent mass at the bottom of a silo is expressed as: 
\begin{equation}
m_{a}=m_{sat}\left[1-exp\left(-\frac{m_{fill}}{m_{sat}}\right)\right],\label{eq:equiv mass}
\end{equation}
where the saturation mass is given by $m_{sat}=\frac{\rho\pi R^{3}}{2K\mu_{s}}$,
$\mu_{s}$ is the Coulomb static friction coefficient between
the grains and the wall, $m_{fill}$ the filling mass, and $R$ the
radius of the silo. In the case of a squared silo of size $L$, we
substitute the radius by the average distance from the center:

\begin{equation}
    {R'=\frac{2L}{\pi}\int_{0}^{\pi/4}\cos\theta\textrm{d}\theta=\frac{2L}{\pi}\ln\left(\sqrt{2}+1\right)\approx0.56 L}
\end{equation}

Choosing a friction $\mu_{s}=0.5$ for PMMA-glass (PolyMethyl MethAcrylate), as measured in
\citet{Cambau2013}, and our estimate of $K=0.2$ (Section \ref{subsec:Data interpretation}),
the saturation mass is $m_{sat}=572$ kg. Since the full box mass,
$M_{fill}=195$ kg, is smaller than the saturation mass, the pressure
at the bottom given by the equivalent mass (eq. ) is $p=M_{a}/S=6.6$
kPa, which it is very close to the hydro-static pressure $p_{min}=7.6$
kPa. Eq. (\ref{eq:equiv mass}) becomes a
depth-dependent ($x$) pressure by considering a constant density
to describe the filling mass: $m_{fill}=x\rho S$: 
\begin{equation}
p\left(x\right)=\frac{m_{sat}}{S}g\left[1-exp\left(-\frac{x\rho S}{m_{sat}}\right)\right].\label{eq:pressure Janssen}
\end{equation}
This pressure profile, only alid when no external
pressure is applied can be used in the numerical simulation. Nevertheless, the difference with a homogeneous model is only significant at the surface. In the geometry of the experiment, at 15 cm depth, the wave speed reaches 93\% of its average in the measuring zone.

\end{document}